%% file: pdf4eg.tex
\RequirePackage{lineno} 
\documentclass[12pt]{article}

\usepackage{amsmath,amssymb,graphicx}
\usepackage{srcltx,cite}
\usepackage{url}

\input pdf4egpre                

\begin{document}


\input front               

\input pdf4eg1                  

\input pdf4eg2                  

\input pdf4eg3                  

\input pdf4app                   

\input pdf4egcit                

\end{document}

%% file: pdf4egpre.tex
\newcommand{\DATE}  {\today}

\newcommand{\PPrtNo}
{MSUHEP-091021, SMU-HEP-09-34
}
\newcommand{\TITLE}
{Parton distributions for event generators}

\newcommand{\AUTHORS}
{Hung-Liang Lai,$^{1}$ Joey Huston,$^{2}$ Stephen Mrenna,$^{3}$
Pavel M. Nadolsky,$^{4}$ \\
Daniel Stump,$^{2}$ Wu-Ki Tung,$^{2,5,\dagger}$
C.-P.\ Yuan$^{2}$}
\newcommand{\INST}
{ $^{1}$Taipei Municipal University of Education, Taipei, Taiwan \\
$^{2}$Department of Physics and Astronomy, Michigan State University,
\\
East Lansing, MI 48824-1116, U.S.A.\\
$^{3}$Fermilab, Batavia, IL 60510 U.S.A. \\
$^{4}$Department of Physics, Southern Methodist University, Dallas, TX \\
$^{5}$Department of Physics, University of Washington, Seattle, WA
98105, U.S.A. \\
$\dagger$ Deceased }

\newcommand{\ABSTRACT}
{In this paper, conventional global QCD analysis is generalized
to produce parton distribution functions (PDFs) optimized for use with event 
generators at the Large Hadron Collider (LHC). 
This optimization is accomplished by complementing
usual constraints on the PDFs from the existing hard-scattering 
experimental data with those needed to reproduce cross sections 
for key scattering processes at the LHC, 
as predicted by the best available theory, in the joint input 
to the global analysis. With the optimized PDFs, 
predictions obtained by event generators at a given order 
in the QCD coupling strength reproduce the representative LHC 
cross sections computed at one higher order. In the present study, 
the optimized PDFs for leading-order event generators were developed.
Several optimization strategies and resulting candidate PDF sets 
(labeled as CT09MCS, CT09MC1 and CT09MC2) are compared 
with those from other approaches.}

%% file: front.tex

\begin{titlepage}

\begin{tabular}{l}
\noindent\DATE
\end{tabular}
\hfill
\begin{tabular}{l}
\PPrtNo
\end{tabular}

\vspace{1cm}

\begin{center}
\renewcommand{\thefootnote}{\fnsymbol{footnote}}
{
\LARGE \TITLE
}

\vspace{1.25cm}
{\large  \AUTHORS}

\vspace{1.25cm}

\INST
\end{center}

\vfill

\ABSTRACT                 

\vfill

\newpage
\end{titlepage}

%% file: pdf4eg1.tex
\section{Introduction
\label{sec:intro}}

Monte Carlo event generators play a critical role in all stages of modern
particle physics, from detector design to calculation of acceptances
and interpretation of experimental results.
A key input needed for event generators is parton distribution functions 
(PDFs). The PDFs are used
(1) in the evaluation of the hard subprocess matrix elements,
(2) in the backward showering algorithm for initial-state radiation, 
and (3) in the calculation of the multiple parton interactions
that make up the bulk of the underlying event.
The latter, in particular, requires extensive tuning which depends strongly
on detailed features of the input PDFs. 
 
A long-standing question, and dilemma, in this regard has been:
what are the appropriate PDF sets that one should use with the
available event generators, in particular, with the most 
mature and widely used leading-order (LO) generators?
For next-to-leading-order (NLO) event generators,
such as MC@NLO~\cite{Frixione:2002ik}
and POWHEG~\cite{Nason:2004rx,Frixione:2007vw},
the  answer is reasonably straightforward:
use NLO PDF sets defined in a compatible factorization scheme.\footnote{%
This factorization scheme must agree with the specific algorithm for
treatment of exclusive final states in the NLO event generator \cite{Collins:2002ey}.}
However, the number of processes implemented in a NLO Monte Carlo
framework is still limited, and the use of LO Monte Carlo programs
is more widespread. But for LO Monte Carlo event generators,
the choice of the PDFs and their order is non-trivial. 

In practice, most applications of LO event generators
have been using available LO PDFs.
Certain alternative practices,
such as using NLO PDFs in LO event generators, have been also proposed 
to address some of the issues.
It has been observed~\cite{Campbell:2006wx}
that a better agreement with fully NLO predictions,
both in terms of the shape and normalization of the cross section,
and of acceptances calculated with experimental cuts,
can often be obtained when LO event generators use NLO PDFs.
But these alternatives have their own known drawbacks,
particularly with the determination of the underlying event,
and are not robust for all processes.\footnote{The PYTHIA8~
\cite{Sjostrand:2007gs} framework allows one to use separate PDFs 
for the generation of the hard-scattering
portion of the event and for the generation of the underlying event.
Thus, one could use a NLO PDF for the matrix element
and a LO PDF for the parton showering/underlying event.
Formally, all parts of an event must be computed with the same PDF set;
but, if the $x$ and $Q^2$ regimes where the two PDF sets are invoked
are very different, the inconsistency should not be too serious.}
The urgent need for better-performing event
generator calculations has stimulated much discussion
at recent conferences and workshops about PDFs 
that are tailor-made specifically for event generators.
This general idea obviously makes sense;
the question is how to construct these special PDFs?

To address this question, it is necessary to distinguish between 
two different sources of mismatches between the conventional LO PDFs 
and their event generator applications.
The first problem is due to intrinsic limitations of the LO global
analysis or LO calculations that make their ``predictions" 
inherently unreliable at
higher energies, beyond that of the input experimental data included in the
global analysis (e.g., at the LHC), and for new physical processes that are
not included in the global analysis (e.g., top quark and Higgs production).
This problem has been discussed in literature~\cite{Campbell:2006wx}.
Suggestions have been made~\cite{ThorneSh} to remedy the known
deficiencies of the LO calculations by relaxing constraints from 
the momentum sum rule and other common practices, basically  by a trial-and-error
approach determined by the {\it a posteriori}  result.
This paper will try to address this problem in a more direct 
way by utilizing the power of the global analysis itself and
by going beyond its conventional method. 

The second mismatch is associated with the initial-state radiation (ISR) 
that is present in the event generators, but not
in the global analysis determining the LO PDFs. 
This problem, in principle, depends on the type of the event generator,
given that each of them handles ISR differently.
The differences in the ISR treatment should formally be taken into account
when deriving the appropriate input PDF sets. In practice, at LO accuracy,
the main impact of the radiation is kinematic in nature,
with further subtleties being formally at NLO 
and thus beyond the scope of this study.
In an initial-state parton shower, gluons are radiated at finite angles.
In the DGLAP formalism used in global PDF fits, gluons are assumed to be 
collinear. Thus, to produce, say, a $W$ or a Higgs boson 
at a particular rapidity, a larger momentum fraction for the incoming partons
is required in a parton shower Monte Carlo program 
than in a fixed-order formalism,
resulting in a kinematic suppression. 
We discuss the size of this suppression in Section~\ref{sec:PS}
and show that the effect, although noticeable, does not significantly affect
predictions at the LHC, in comparison to more pronounced differences arising from the choice between the LO and NLO PDFs. 

%% file: pdf4eg2.tex
\section{Global analysis of PDFs For LO event generators
\label{sec:GaPdf}}

Conventional global analyses determine PDFs by fitting theoretical QCD
cross sections to the existing hard scattering data.
Universality of PDFs and their calculable QCD evolution to
higher scales allow us then to make predictions at higher energies 
and for new processes. For most applications, 
especially those relevant for event generators,
this principle works well at NLO, since the accuracy of perturbative 
QCD predictions at this order usually matches the current 
and expected experimental precision.\footnote{%
For some processes, theoretical errors are exceptionally large;
then even higher-order terms beyond NLO are needed.}
But when PDFs are determined in a LO global analysis, 
using existing experimental data, they are known to have incorrect
behavior both at small and large partonic momentum fractions $x$, 
due to missing large terms that first arise in the hard matrix
elements at a higher order (NLO). Many  Tevatron/LHC cross sections 
tend to be larger at NLO than at LO for commonly used scales,
i.e., the K-factor (the ratio of the NLO to LO cross sections) 
tends to be larger than 1 -- see, for example,
Ref.~\cite{Campbell:2006wx} and
Table~\ref{tab:K-fact} later in this paper. As a consequence, 
when conventional LO PDFs are used in LO generators,
predictions at  high energies (such as at the LHC or, in some cases,
the Tevatron) and for new physical processes can be quite unreliable,
both in magnitude and shape \cite{Campbell:2006wx,ThorneSh}. 
Alternative prescriptions, such as using LO matrix elements 
with NLO PDF sets, may better reproduce the shape of the full NLO
cross section; but, as already mentioned, other issues with the
normalization and underlying event still remain~\cite{Campbell:2006wx}. 

Since, by definition, we are constrained to use LO matrix elements
in LO generators, this long-standing dilemma
can be resolved---to the extent possible---only by trying
to optimize factorization of LO cross sections, i.e., 
by finding better PDFs and possibly more sensible
renormalization and factorization scales for each LO cross section. 
This can be carried out most systematically by redefining the goal and 
strategy of the global QCD analysis.\footnote{%
The need to rethink the strategy of the global analysis for event
generators is also implicit in other attempts  \cite{ThorneSh, pdf4mc}
 to address the same problem.}
 In a conventional global analysis, the PDFs are optimized to fit
\emph{the existing experimental data}.
For event generator applications, this is not the main purpose;
rather, it is equally, if not more, important to
\emph{produce reliable predictions} at higher
energies and for new processes.
In fact, the efficacy of a PDF set for event generator applications,
particularly LO ones, is mostly judged by how its predictions for future
colliders meet expectations.\footnote{%
And it is mainly on this ground that the conventional LO PDFs have
been deemed unsatisfactory.}
But, prior to having real data at these colliders,
what constitutes the correct expectations?
This is where the existing NLO and NNLO calculations come in.
There is a good reason to believe that, for standard model (SM) processes,
the predictions of QCD at NLO and NNLO orders will be reasonably reliable.
They can be used as a sensible substitute
for nature (or ``truth" as called in Ref.\,\cite{ThorneSh}).

This observation immediately suggests that the most direct, and effective,
way to obtain PDF sets for event generators is to generalize the
conventional global QCD analysis to utilize the best estimates of key
physical processes at future colliders (to ensure reliable predictions),
\emph{in parallel with} the existing experimental data sets
(to ensure reasonable agreement with nature at currently
available energy scales),
\emph{as joint inputs to the global fitting}.
In principle, this idea can be applied at both LO and NLO;
however it is only of practical interest for LO event generators
at present.\footnote{%
To improve NLO PDFs for use with NLO event generators, one could
supplement existing experimental data input by predicted high energy
cross sections calculated in NNLO, whenever these calculations are
available.}
For this purpose, we can implement the constraints of
``nature" at high energies in the form of pseudodata sets 
generated by NLO calculations for representative physical processes 
that are sensitive to various flavors of partons: light quarks, gluons, and heavy quarks.

Even if this basic idea of a global analysis of PDFs optimized for event
generators is quite simple and natural, a few relevant considerations need
to be pointed out before going into details.
First, since we are focusing on PDFs for LO generators,
we must use LO matrix elements in the calculations for the global fitting.
But we know already that LO matrix elements provide only the most basic
approximations to the true theory; therefore, even the most optimized PDFs
cannot be expected to fit well both the lower-energy experimental data,
especially in deep-inelastic scattering processes, 
and the higher energy pseudodata at the same time.
These PDFs represent the \emph{best compromise} that can be obtained within
the restrictions of the LO matrix element approximation.
They are intended solely as an input to LO event generators
for predictions {\it at the LHC}, with an eye on their inherent limitations. 

In its typical application, a LO event generator is not used to predict
absolute cross sections {\it per se}, but rather 
to calculate detector acceptances for, and backgrounds to,
physics processes of interest, 
in conjunction with detailed detector simulations.
It is desirable that the LO event generators produce reasonably
accurate normalizations for the cross sections,
although it is understood that higher-order contributions 
not included in the LO generators may introduce sizeable corrections.
But it is even more important that kinematic shapes, 
such as rapidity distributions, be accurately described, so that 
event acceptance derived from these distributions is close to reality. 

It should be readily emphasized that the PDFs generated in
this modified manner are not ``leading-order PDFs"---
rather, they are ``PDFs for leading-order Monte-Carlo programs'', 
or ``LO-MC PDFs''.
This distinction needs to be made,
since there are still lingering misconceptions about the need
to ``match orders" in literature and in public discussions.
Event generators, including ``LO event generators",
have some elements of higher-order contributions
and, in this sense, are not at the stated order in the QCD coupling.
Because of these two considerations, the global fits we are
performing have the freedom of choice on several fronts,
all of which can impact the numerical results.

$\star$ \textbf{The order of $\alpha_s$:}
We have the choice of using either a 1-loop or 2-loop version of the QCD
coupling $\alpha_s$.
Nominally, a 1-loop $\alpha_s$ may be considered as more appropriate
with a LO event generator, but some parton showering models  
prefer a 2-loop $\alpha_s$.
We also have the freedom to set $\alpha_s$ free in the global fit
or to tie it to the world average.
We choose to fix $\alpha_s(M_Z)$ at the world average
(0.118 at two loops and 0.130 at one loop),
for convenience and compatibility with the previous CTEQ PDF sets.

$\star$ \textbf{Factorization scales:}
Should the renormalization and factorization scales 
for different processes be fixed,
or could they be discretionally chosen, or even be fitted,
in order to get the best agreement?
The motivation for considering flexibility here is because LO calculations
are notoriously scale-dependent for most processes.
Changes in the scales can affect both the normalization
and the shapes of the Tevatron and LHC cross sections. But this
flexibility can be also employed to advantage, by finding the LO scale values
that provide the best approximation to the
real data and NLO pseudodata.\footnote{The scale dependence is usually
moderated at NLO. In our global fits  at NLO, the scales are always
fixed at their nominal values.}

$\star$ \textbf{Momentum sum rule:}
This sum rule relates PDFs or different flavors in order to conserve the
total momentum carried by the partons. 
Can it be relaxed in this kind of global analysis?
This possibility was brought forth by earlier attempts to fix the problems
of existing LO PDFs~\cite{ThorneSh} by putting more gluons 
in the high-$x$ region than otherwise would be allowed in a LO fit.
Is this still needed in our approach,
which automatically puts more partons into the relevant $x$ region 
because of the constraints imposed by the NLO pseudodata?
Even when relaxation  of the momentum sum rule is not required,
could it still improve the results?
We shall answer these questions by performing parallel global
analyses with and without enforcing the momentum sum rule
and by comparing their outcomes.

$\star$ \textbf{Selection and construction of NLO pseudodata sets
to represent ``nature" at high energies:}
There is clearly a great deal of latitude in doing this.
The selection of physical cross sections to be represented
in these theoretical data sets
is guided by the importance of the process for the LHC physics,
and by the parton flavors that these processes are sensitive to.
One would like to ensure that all parton flavors, in most ranges of $x$,
are covered by these constraints.
Since these pseudodata sets are used in a fitting procedure,
one must also assign ``errors" to each data point, 
as well as overall weights of the $\chi^2$ values 
contributed by each pseudodata set.
These can be guided by the estimated theoretical and expected
experimental errors, but are ultimately subjective.

To satisfy phenomenological considerations, 
the LO-MC PDFs 
should 
\begin{itemize}
\item{behave similarly to the usual LO PDFs 
as $x\rightarrow 0$ (assumed by the current models for the underlying event)
and to NLO PDFs as $x\rightarrow 1$;}
\item{describe the underlying event at the Tevatron
(with a Monte-Carlo tune similar to what is currently used) and extrapolate
to a reasonable level of underlying event at the LHC.}
\end{itemize}

Th NLO pseudodata for the LHC scattering processes 
included in the fit is chosen so as to enforce this desired
behavior of the LO-MC PDFs. As such, we use the single-inclusive $W^+,W^-$  and
$Z^0$ rapidity distributions (affecting the low-$x$ and high-$x$ 
quark distributions), the $b\bar{b}$~\footnote{%
Here, we consider $b\bar{b}$ production only through $gg$ fusion
(with the $b$ quark mass set equal to 4.75 GeV), to constrain the gluon PDF
in the low-$x$ range typical for the underlying event in 
hard scattering collisions. We fit the $b\bar{b}$ mass range from 10-100 $GeV/c^2$.  Hereafter, we refer to this process as $b'\overline{b'}$ since it refers to a restricted set of production subprocesses.}
and $t\bar{t}$ invariant mass distributions,
and the rapidity distribution for a 120 GeV standard model Higgs boson
produced through $gg$ fusion 
(affecting the low-$x$ and high-$x$ gluon distribution).
All NLO pseudodata cross sections were computed using the MCFM 
program~\cite{Campbell:2000bg} and CTEQ6.6M PDFs~\cite{Nadolsky:2008zw}.

When generating the vector boson and the Higgs boson NLO pseudodata,
we have set the renormalization/factorization scale to be equal
to the (pole) mass of the respective boson.
For the scale in 
$t\bar{t}$ production, we have used the  top quark mass (172 GeV),
and, for the scale in 
$b\bar{b}$ production, we have used the invariant mass of the
quark-antiquark pair. All pseudodata cross sections were computed at 14 TeV,
the nominal center-of-mass energy of the LHC.
After the fit, we also checked the level of agreement 
between the NLO predictions
and the LO-MC predictions at 7 TeV and 10 TeV,
the initial running energies of the LHC. 

\begin{figure}
\includegraphics[width=1.0\columnwidth,height=2\textheight,keepaspectratio]
{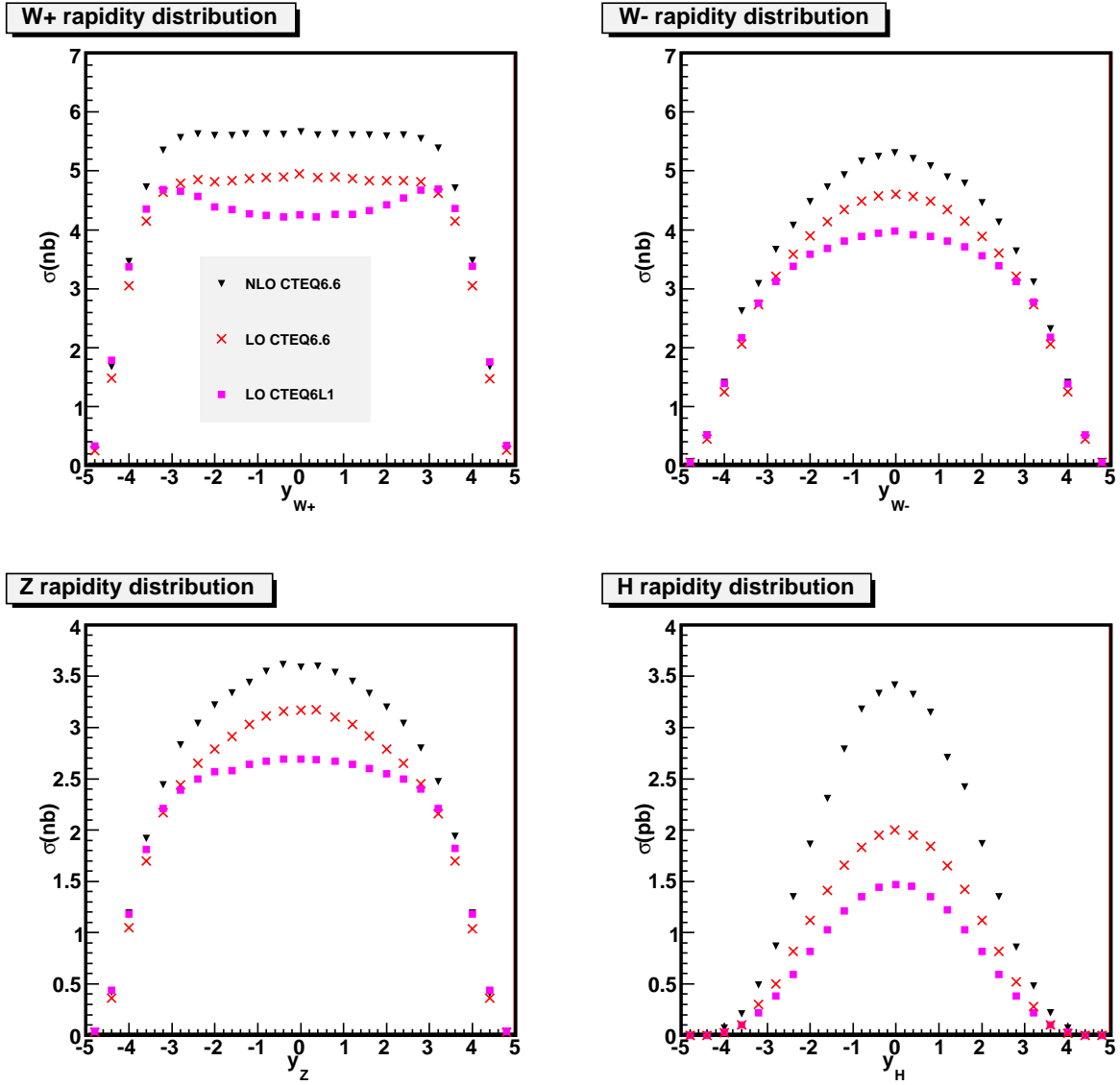}
\caption{A comparison of the NLO pseudodata for SM boson rapidity
distributions (in $\Delta y$=0.4 bins) predicted
at the LHC (14 TeV) to the respective LO predictions based on CTEQ6.6M 
and CTEQ6L1 PDFs. 
\label{fig:Pseudodata-Distribution}}
\end{figure}

To illustrate the scale of the problem we are trying to
address, Fig.~\ref{fig:Pseudodata-Distribution} shows rapidity
distributions for inclusive $W^\pm$, $Z^0$, and Higgs boson production at
$\sqrt{s}=14$~TeV, the key LHC processes. They are computed 
by the MCFM program at NLO using the CTEQ6.6M NLO PDFs, and at
LO using the LO CTEQ6L1~\cite{Pumplin:2002vw} and NLO CTEQ6.6M PDFs, with
the same scale choices. 

As expected, the average normalization of the cross section with the LO hard
part is smaller than that at NLO, regardless of whether the LO PDFs or
NLO PDFs are used. In addition, comparison of the ``LO-LO'' CTEQ6L1
and ``LO-NLO'' CTEQ6.6M distributions reveals significant differences
in the shapes, obviously caused by the input PDFs 
and not by the different orders of
the hard matrix elements. While such differences 
are observed in all four processes, 
the $W^+$ rapidity distribution provides a particularly eye-catching example 
of the danger of using the conventional
LO PDFs with an LO hard cross section (or LO event generators). 
The strong forward-backward 
peaking of the ``LO-LO'' CTEQ6L1 $W^+$ rapidity distribution disappears when
the NLO CTEQ6.6M PDFs are used with the LO hard part.\footnote{%
This peaking 
is caused by the increased magnitude of the CTEQ6L1 $u$-quark distribution at
large $x$, as compared to its CTEQ6.6M counterpart. The same large-$x$ enhancement of
the LO $u$ quarks leads to anomalously large predictions
for ultra-heavy $t\bar{t}$ pair production at the Tevatron.} 
The acceptance for $W^+ \rightarrow e^+ \nu$, computed by a LO event
generator for standard analysis cuts, 
differs when the NLO CTEQ6.6 PDFs are used instead of CTEQ6L1. 
It is thus a misconception that
the strong forward-backward peaking observed in the prediction
based on CTEQ6L1 is a {\it benchmark feature} 
of inclusive $W^+$ rapidity distribution at the LHC.  
In reality, it is primarily an artifact due to inadequacies
of the conventional LO fitting formalism.

The disagreements with the NLO benchmark cross sections are greatly
reduced when the LO cross sections are computed using our LO-MC
PDFs, as will be shown in Section \ref{sec:Results}.  
It is also obvious, from the above description,
that there is a wide range of possible ways
to implement our general approach.

\section{Impact of parton showering
\label{sec:PS}}

The LO-MC PDFs in our study are constructed using fixed-order
(sometimes called ``parton-level'') QCD calculations. 
In practice, these PDFs will be used with LO matrix elements
embedded into a parton shower framework. 
According to initial-state radiation algorithms, shower partons 
are emitted at non-zero angles with finite 
transverse momentum, and not with a zero $k_T$ implicit in the collinear
approximation. It might be argued that the resulting kinematic suppression
due to parton showering (handled differently by various event generators)
should be taken into account when deriving PDFs
for explicit use in Monte Carlo programs.\footnote{%
This was first pointed out to us by Hannes Jung.}

\begin{figure}[t]
\begin{center}
\includegraphics[width=7cm,angle=0]{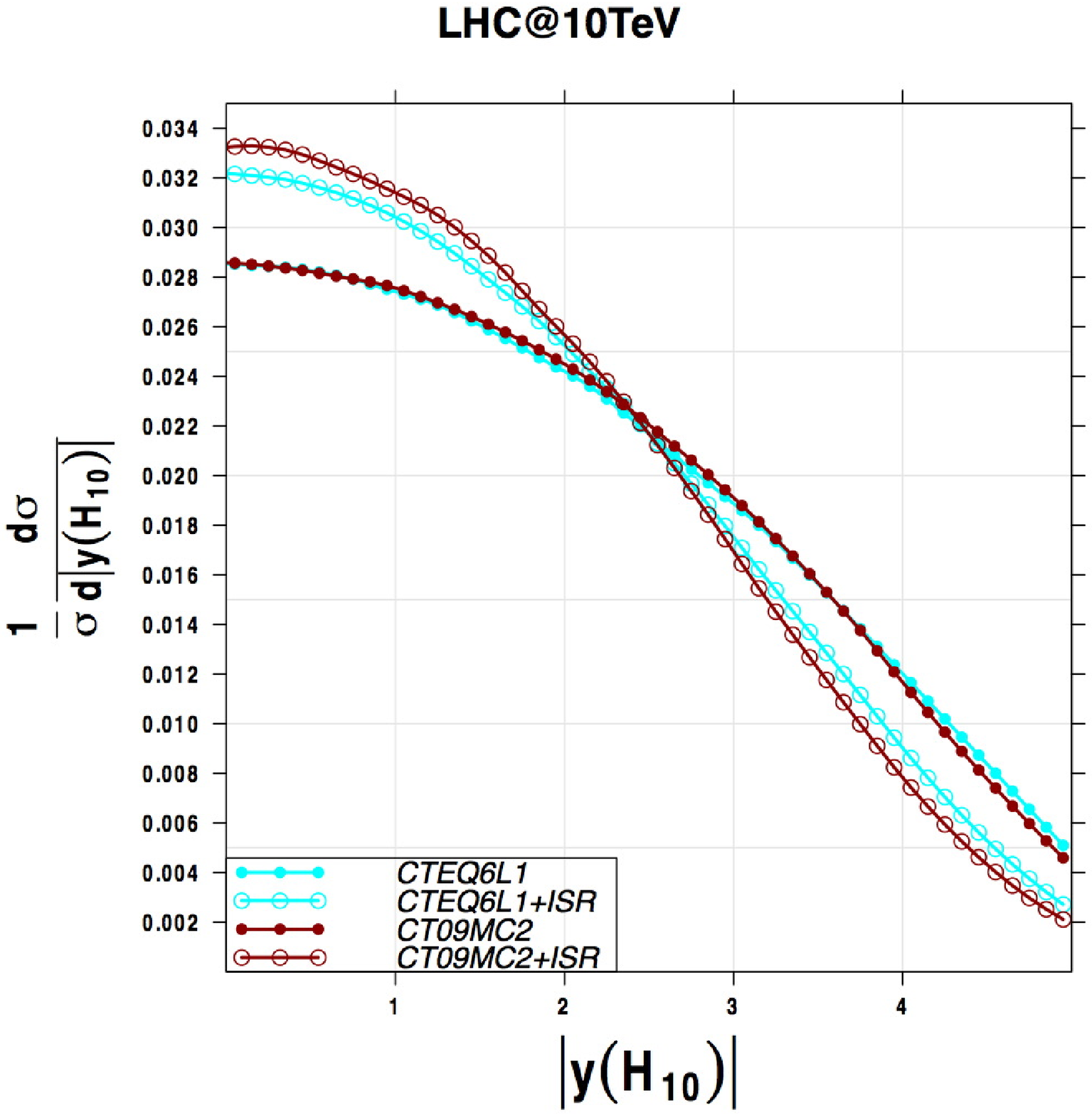}
\includegraphics[width=7cm,angle=0]{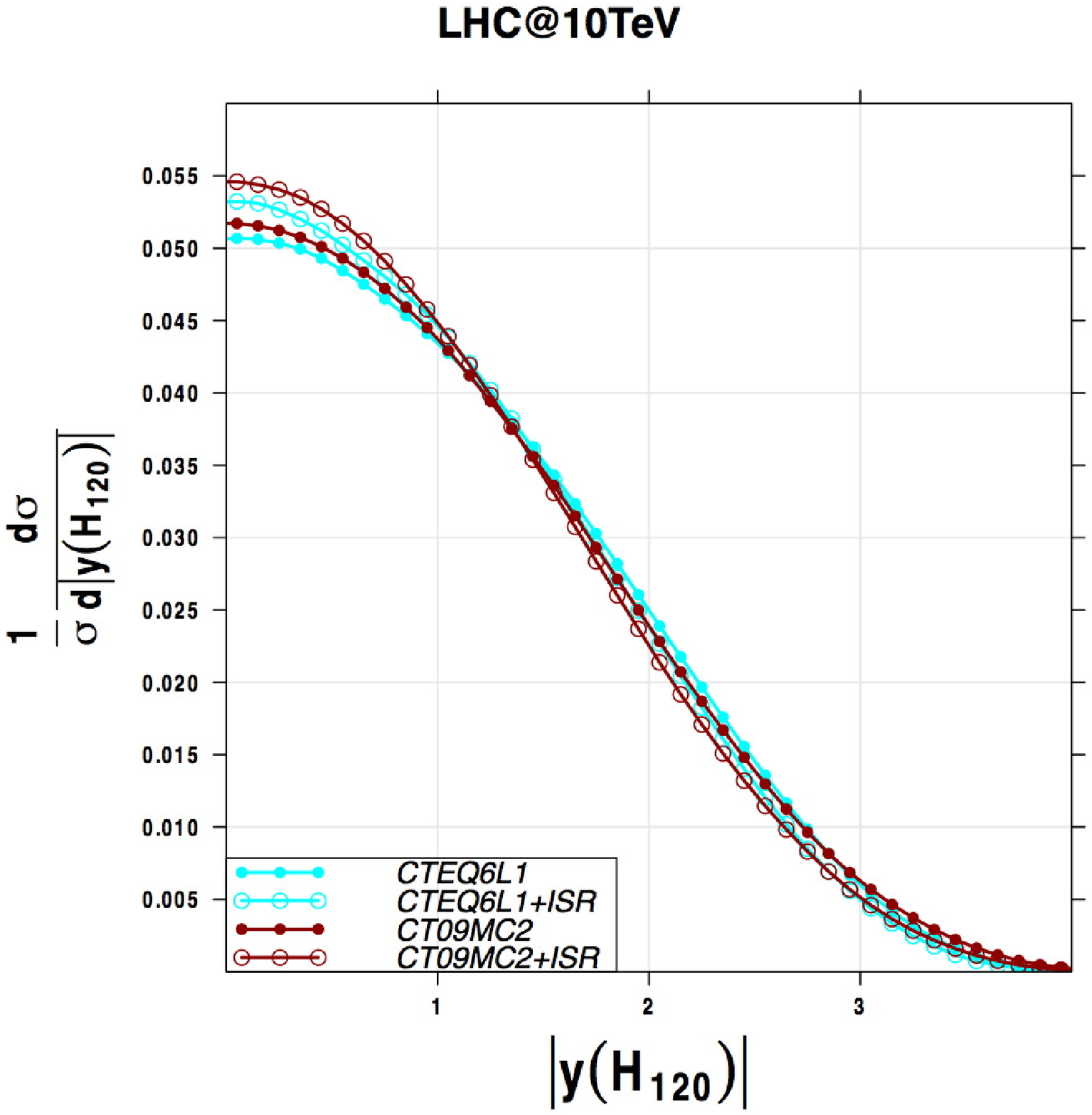}
\includegraphics[width=7cm,angle=0]{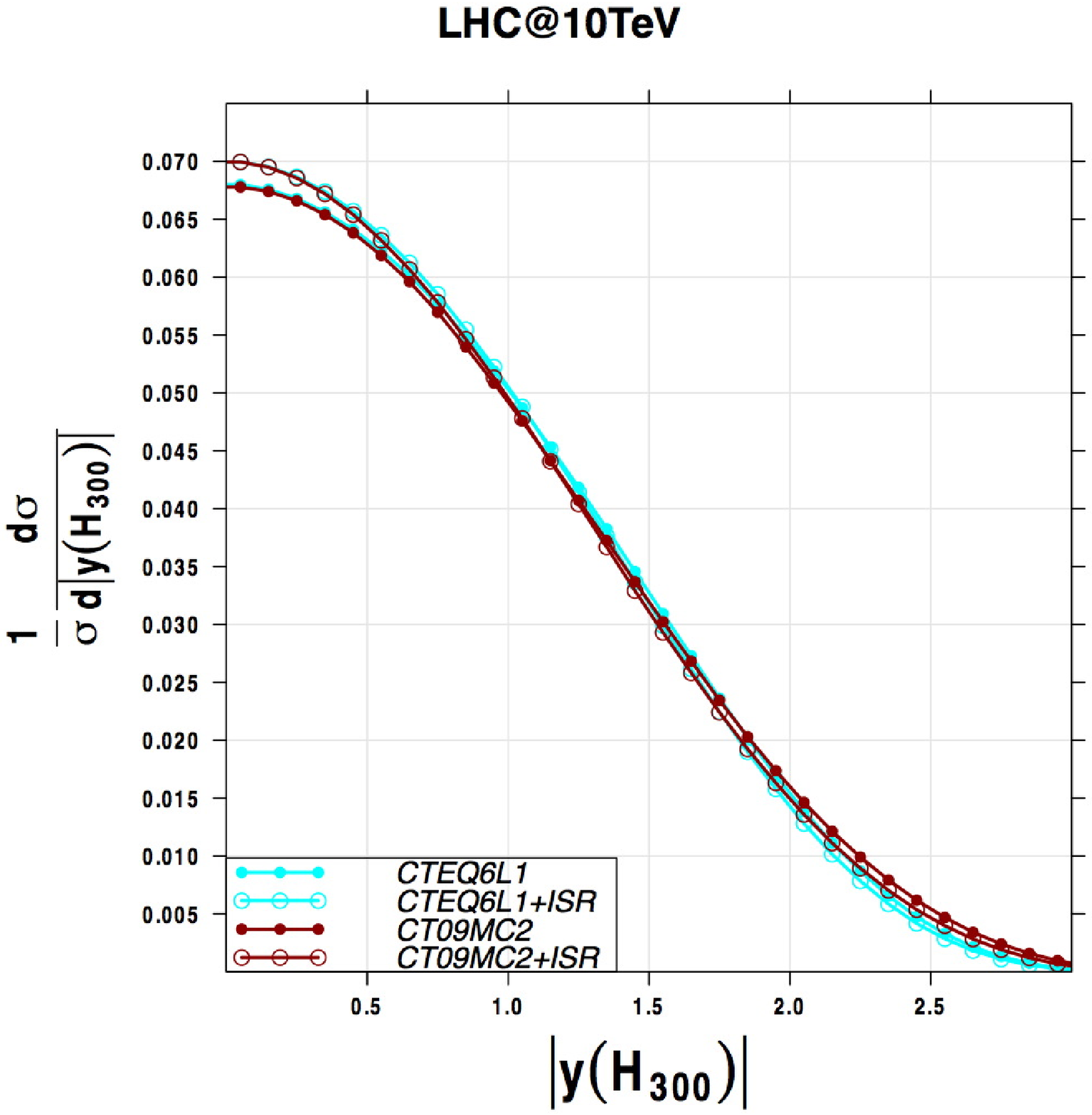}
\end{center}
\vspace*{-0.5cm}
\caption{PYTHIA predictions for production
of a 10 GeV Higgs boson (top left), 
a 120 GeV Higgs boson (top right), and a 300 GeV Higgs boson (bottom) via 
the $gg \rightarrow H$ process at the LHC (at $\sqrt{s}=$10 TeV),
with and without contributions from the initial-state radiation. 
Distributions in the absolute value of the Higgs boson's rapidity $|y|$ are shown.
\label{fig:higgs_ps}}
\end{figure}

To quantify kinematical dependence of this suppression, 
Fig.~\ref{fig:higgs_ps} examines several leading-order 
rapidity ($y$) distributions for SM Higgs boson production via $gg$ fusion,
obtained in the PYTHIA event generator \cite{Sjostrand:2007gs}. 
We compare cross sections
with and without initial-state radiation (ISR) contributions, for 
either the CTEQ6L1 PDFs or one of our new LO-MC PDF sets
CT09MC2 (to be described later). In the top left figure, distributions
for a (toy) 10 GeV mass Higgs boson at the LHC energy $\sqrt{s}=$10 TeV 
are considered.
A sizeable kinematic suppression in the presence of ISR 
is evident at forward rapidities, while the total cross section 
(integrated over the whole rapidity range) 
remains largely unaffected.
These features force the rapidity distribution of such an ultra-light
Higgs boson to be more central with the initial-state parton showering on
than without it. 

In production of a heavier Higgs boson with mass 120 GeV (top right figure),
the effects of the kinematic suppression at forward rapidities
are still evident, but reduced in magnitude. 
For Higgs boson with mass 300 GeV (bottom figure),
the effects of the kinematic suppression are reduced still further. 
This behavior indicates that parton showering 
is not likely to affect greatly the rapidity distributions 
for large-mass phenomena at the LHC, such as for example, 
$t\bar{t}$ production. 

\begin{figure}[t]
\begin{center}
\includegraphics[width=10cm,angle=0]{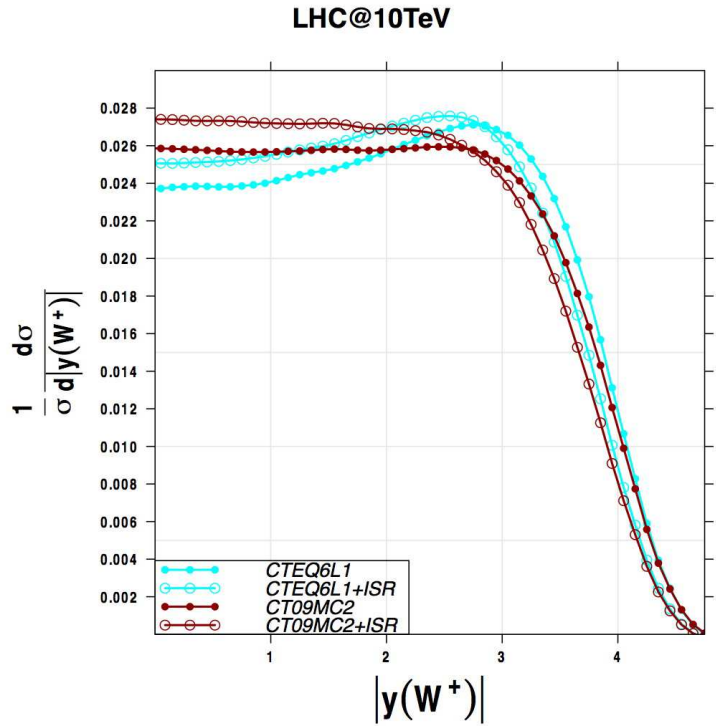}
\end{center}
\vspace*{-0.5cm}
\caption{PYTHIA predictions for rapidity distributions of a $W^+$ boson 
produced via $q' {\bar q} \rightarrow W^+$ process at the LHC (at
$\sqrt{s}=$10 TeV),
computed with CTEQ6L1 PDFs and CT09MC2 PDFs, with and without contributions from the 
initial-state radiation.   
\label{fig:wp_ps}}
\end{figure}

A comparison of PYTHIA predictions for production of a
$W^+$ boson at the LHC (10 TeV) with and without parton showering
is shown in Fig.~\ref{fig:wp_ps}.
For both CTEQ6L1 and CT09MC2 PDFs, 
alterations in the shape of the rapidity distribution caused by 
the parton showering are relatively small. 
In particular, it can be noted that the differences in the shape of
$W^+$ rapidity distributions introduced by conventional LO PDFs (such as CTEQ6L1),
as compared to the NLO cross section, are largely unaffected by the parton
showering. The choice of the PDF set evidently outweighs the
impact of parton showering in the case of $W$ boson production. 

In general, the use of the LO-MC PDFs shifts the production of gauge
bosons to more central values of rapidity. A similar shift occurs 
because of parton showering, but the magnitude of the shift decreases 
as the mass of the final state increases.  The impact of the showering 
also decreases for higher center-of-mass energies (14 TeV, for
example, as compared to 10 TeV). 
For the rest of the paper, unless noted, we will use fixed-order predictions,
although extensive comparisons have been made with parton-shower
predictions as well. 

\clearpage

%% file: pdf4eg3.tex
\section{Results of the modified PDF analysis
\label{sec:ResultsModPDFanalysis}}

\subsection{General considerations
\label{sec:considerations}}

The LO-MC PDFs (designated as CT09MC PDFs) 
are constrained by including the same 
existing experimental data sets as those used 
in the CTEQ6.6 PDF analysis~\cite{Nadolsky:2008zw}, combined with the
pseudodata on NLO cross sections for five representative LHC
scattering processes discussed in Sec.~\ref{sec:GaPdf}.
Correlated systematic error information
is used for all experimental data sets.

To give an idea about the impact the NLO radiative contributions, 
a fully NLO global fit in the CTEQ framework, {\it with no pseudodata},
results in a $\chi^2/\mbox{d.o.f.}$ close to 1 for a sample
of around 2700 data points.
If the fit is carried out instead at LO, with a 1-loop $\alpha_s$,
the $\chi^2$ worsens by about 30\%.
If $\alpha_s$ is evaluated at two loops,
the $\chi^2$ is larger than that at NLO by 20\%;
i.e., the 2-loop $\alpha_s$ experssion improves $\chi^2$ in the
LO fit by about 10\%. Thus, the data prefer more rapid variation 
of $\alpha_s$ with $Q^2$ provided by its two-loop expression.

If, in addition, the momentum sum rule is relaxed,  
modest improvements in the global $\chi^2$ are observed, accompanied
by a violation of the momentum fraction sum on the order of 3\%.
Allowing more gluon momentum does improve the LO-MC fit to some of the
(regular) data sets, but results in a worse fit to other data sets.
Thus, we find it difficult to achieve as small a value of $\chi^2$
in the LO-MC fits as in the NLO fit,
even when the momentum sum rule is relaxed. 

\subsection{Numerical results
\label{sec:Results}}

\subsubsection{CT09MCS PDFs} 
We  now consider the LO-MC PDFs produced with the NLO pseudodata
included in our data set. First, we consider the case where
the momentum sum rule is kept intact,
but the factorization scales in the LO matrix elements corresponding
to the pseudodata are allowed to vary as free parameters. The normalization of 
the LO calculation for each pseudodata set $i$
is also allowed to float to reach the best agreement with the NLO
cross section, which is equivalently described by a floating 
normalization of each pseudodata set, denoted by $N_i$.
The effective K-factor (NLO/LO) for the pseudodata 
is then given by $K_i = 1/N_i$. We will name the LO-MC PDF set
resulting from this approach as ``CT09MCS'', where {\it S} signifies the
varied {\it scales} in the fit to the pseudodata.

In practical terms, the factorization scale $\mu_i$
for each pseudodata set, taken to be the same as the
renormalization scale, is allowed  to vary within
a factor of four around the nominal scale
defined for each process in Section~\ref{sec:GaPdf}.
A $\chi^2$ penalty is assigned for deviations of the normalization
$N_i$ from unity, and the weights applied to $\chi^2$ values from 
the pseudodata sets can be varied as well.
For this  exercise, 
we use only the 2-loop $\alpha_s(m_Z)$.
As stated previously, the value of the (2-loop) $\alpha_s(m_Z)$
is fixed at the value of 0.118 used in the CTEQ6.6 global fit. 

\begin{figure}
\begin{center}
\includegraphics[width=0.6\columnwidth,height=1.2\textheight,keepaspectratio]
{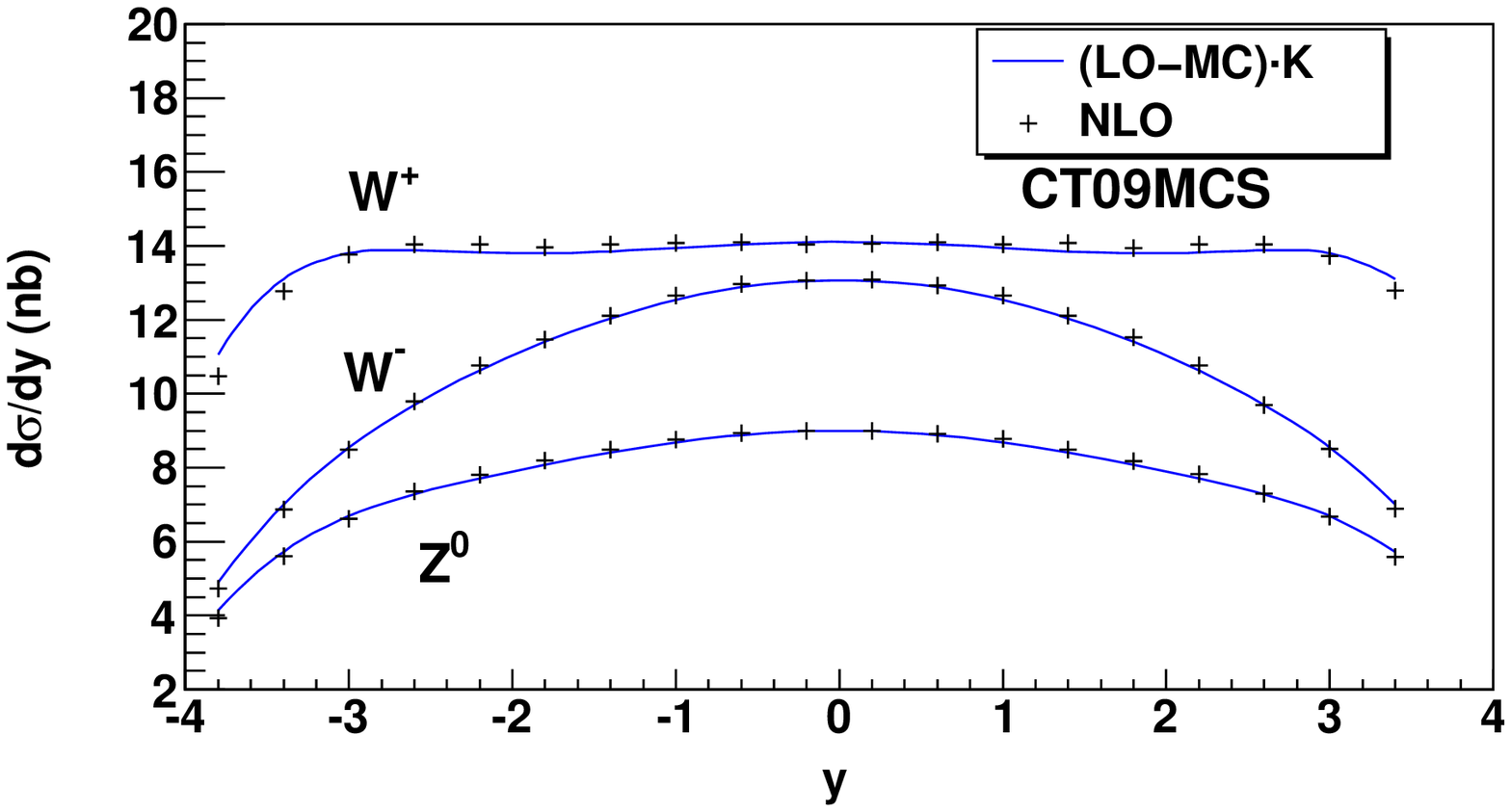}
\includegraphics[width=0.6\columnwidth,height=1.2\textheight,keepaspectratio]
{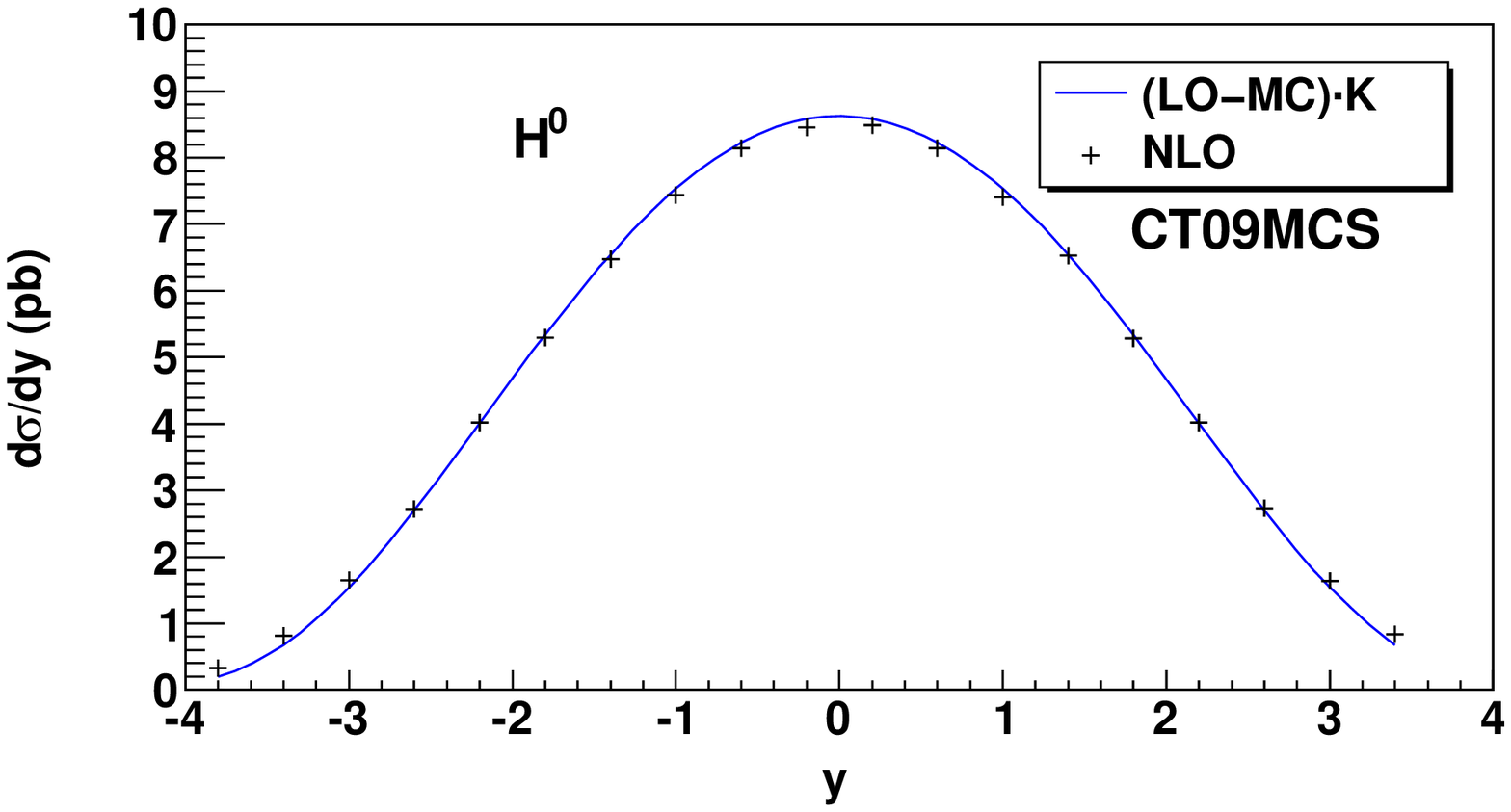}
\end{center}
\caption{Comparison of the NLO pseudodata cross sections for $W$, $Z$
and Higgs production at the LHC (14 TeV) with the LO predictions using
CT09MCS PDFs. The scale choices and effective K-factors applied 
to the LO-MC cross sections 
are listed in Table~\ref{tab:fitted-Norm-scale}.
\label{fig:WZ_cp}}
\end{figure}

\begin{figure}[ht]
\begin{center}
\includegraphics[width=0.6\columnwidth,height=1.2\textheight,keepaspectratio]
{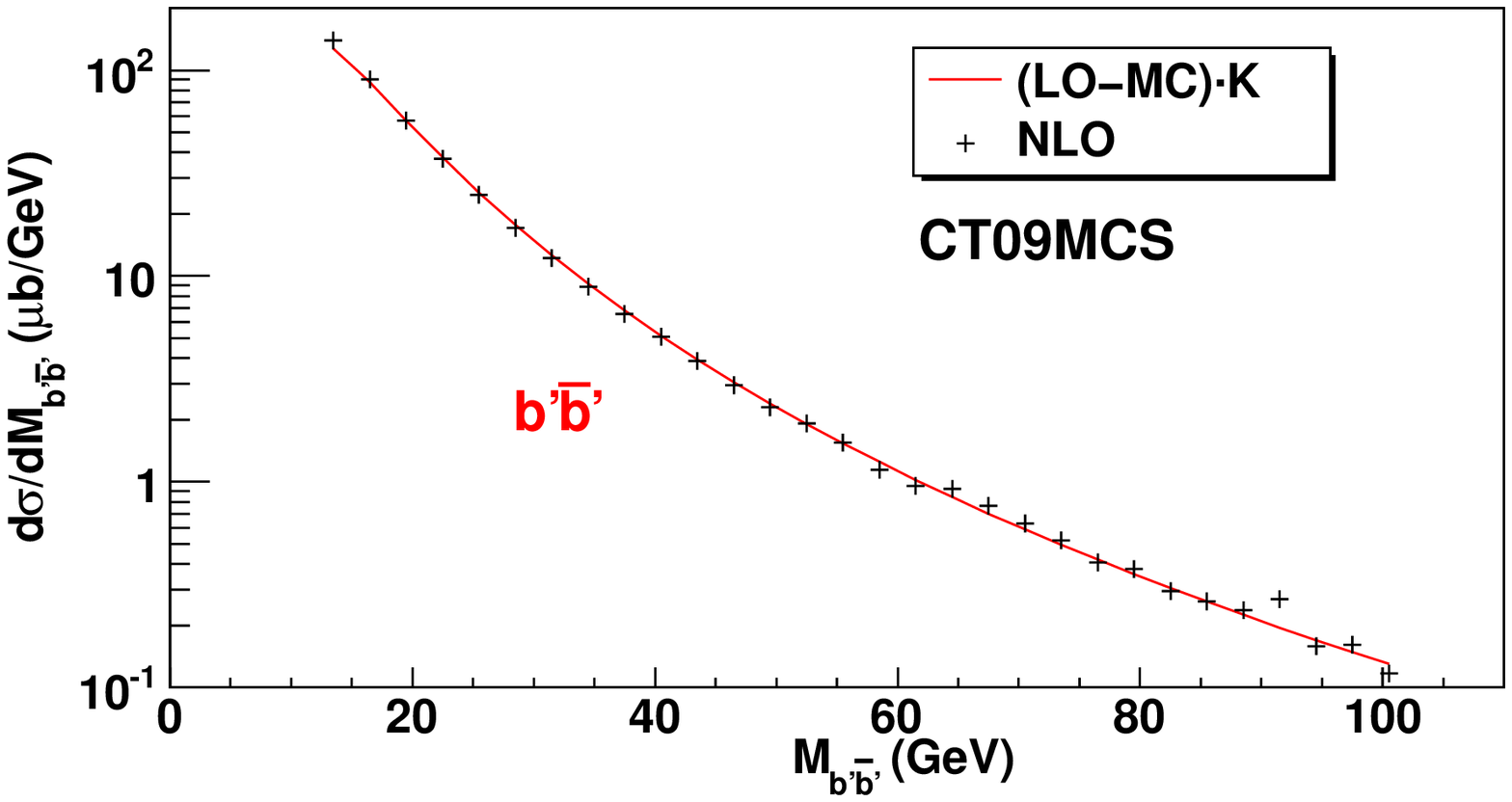}
\includegraphics[width=0.6\columnwidth,height=1.2\textheight,keepaspectratio]
{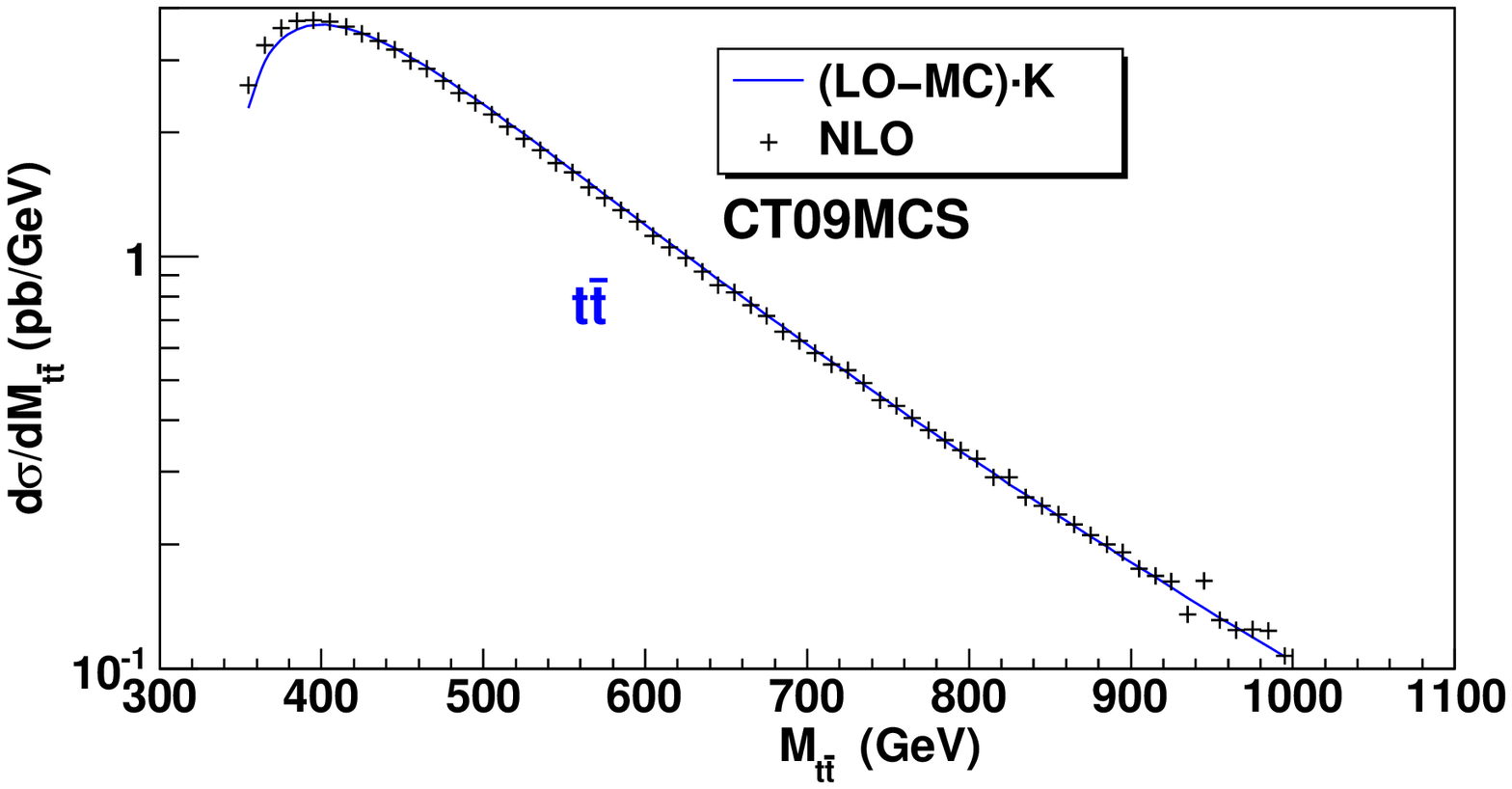}
\end{center}
\caption{Comparison of the NLO pseudodata cross sections
for $b'\overline{b'}$ and $t\bar{t}$ production
at the LHC (14 TeV) with the LO predictions
using CT09MCS PDFs.
The scale choices and effective K-factors applied 
to the LO-MC cross sections
are listed in Table~\ref{tab:fitted-Norm-scale}.
\label{fig:bb_cp}}
\end{figure}

In the CT09MCS approach, the optimum $\chi^2$ is obtained with the scales given in
Table~\ref{tab:fitted-Norm-scale},
with each scale being within a factor of 2 or so from the nominal value.
A comparison of the CT09MCS predictions with the NLO pseudodata is presented
in Figs.~\ref{fig:WZ_cp} and \ref{fig:bb_cp}. The NLO cross sections
are shown with their true normalization, while the LO-MC cross
sections are multiplied by the best-fit K-factors
listed in Table~\ref{tab:fitted-Norm-scale}. Excellent agreement between the CT09MCS and NLO cross sections  is observed for all four scattering processes.

\begin{table}
\begin{center}
\begin{tabular}{|c|c|c|c|c|c|c|}
\hline 
  & $W^{+}$  & $W^{-}$  & $Z$  & $H$  & $t\bar{t}$  & $b'\overline{b'}$
\tabularnewline
\hline
\hline
$\mu_{i}$  & 1.96 $M_{W}$  & 1.96 $M_{W}$  & 1.96 $M_{Z}$  &
 1.06 $M_{H}$  & 1.41 $M_{t}$  & 0.40 $M_{b'\overline{b'}}$
\tabularnewline
\hline 
$K_{i}$  & 1.11  & 1.09  & 1.09  & 1.87  & 2.09  & 4.09
\tabularnewline
\hline
\end{tabular}
\caption{The fitted $\mu_{i}$ and $K_{i}$ for each pseudodata set, obtained
using CT09MCS PDFs. As a reminder, $b'\overline{b'}$ refers to production only through the $gg$ sub-process. 
\label{tab:fitted-Norm-scale}}
\end{center}
\end{table}

\subsubsection{CT09MC1 and CT09MC2 PDFs} 
In the second approach, we again fit the real experimental 
data and NLO pseudodata together, 
but relax the momentum sum rule
and fix the factorization scales at their nominal values. The pseudo-data normalizations are allowed to float, as before. 
We obtain two PDF sets, designated as  CT09MC1 and CT09MC2,
determined with the 1-loop and 2-loop  expressions for $\alpha_s$,
respectively.  In this approach, good agreement with the NLO 
pseudodata is reached only at the expense of a worse agreement with
the real data. We balance between describing the real
data and LHC pseudodata by assigning an extra weight to the
pseudodata to  better reproduce the pseudodata's normalization and shape.
As the weight of the pseudodata in the global fit is increased,  
(i) the pseudodata normalizations get closer to unity,
(ii) larger violation of the momentum sum rule is observed,
(iii) the quality of agreement with the real data sets deteriorates 
progressively, with $\chi^2$ values for the real data being 
worse by 10-20\% for the CT09MC1 and CT09MC2 fits than without the
pseudodata. 
The 2-loop $\alpha_s$ expression 
results in slightly lower normalizations
$N_i$ for the pseudodata sets and a slightly larger violation of the
momentum sum rule than in the case of the 1-loop $\alpha_s$,
but in a similar level of agreement with the real data set. 

The final CT09MC1 and CT09MC2 PDFs thus 
present a compromise that tries for a better
shape and normalization for the pseudodata without sacrificing 
reasonable (LO) description of the real (non-LHC) data sets.

\subsubsection{CT09MC2 predictions for selected LHC cross sections}
\begin{figure}[ht]
\includegraphics[width=1.0\columnwidth,height=2\textheight,keepaspectratio]
{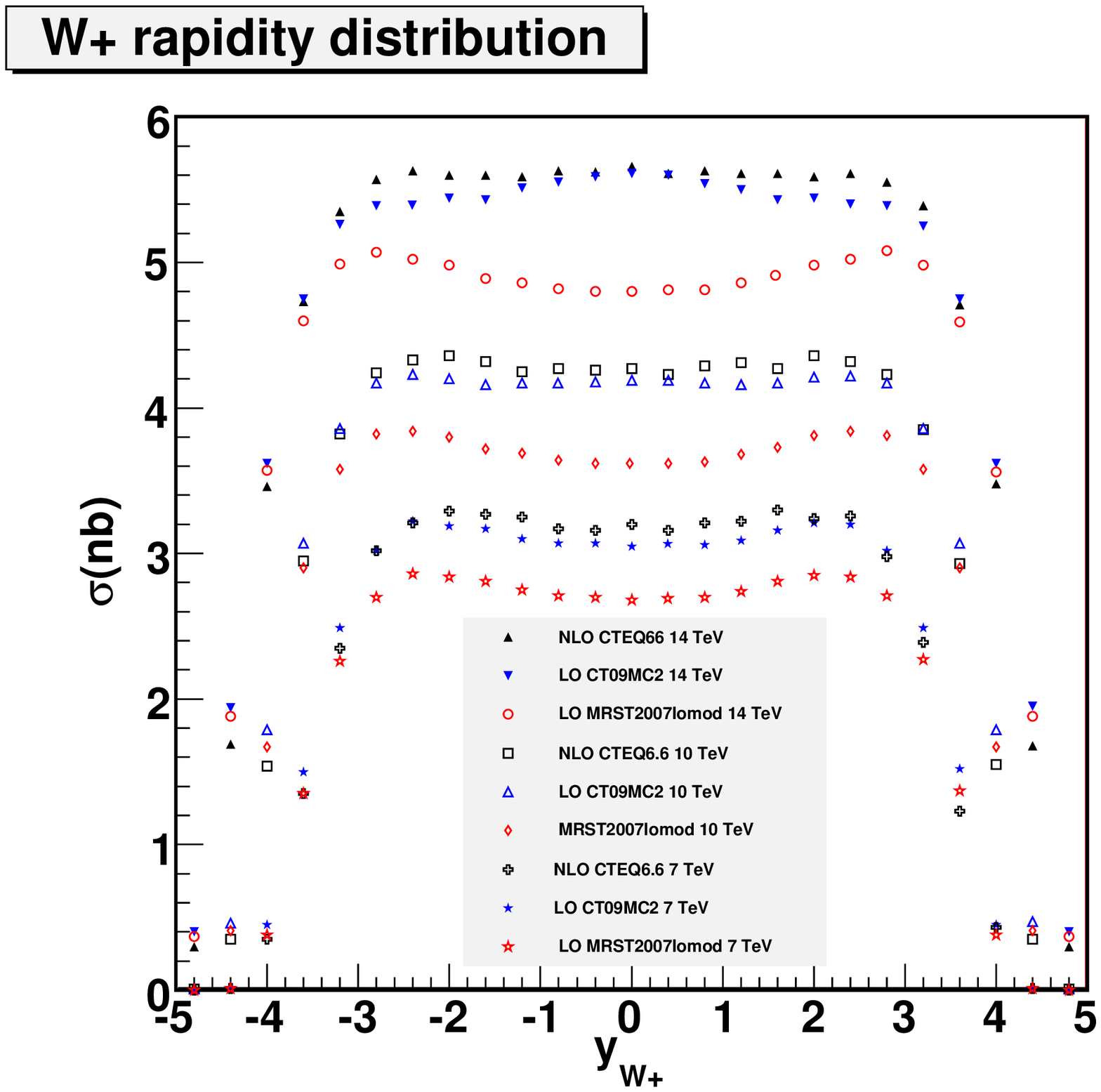}
\caption{Predictions for the $W^+$ rapidity distribution at the LHC
($\sqrt{s}=$7, 10 and 14 TeV) in $\Delta y$ =0.4 bins,
given at NLO using the CTEQ6.6M PDFs,
and at LO using the CT09MC2 and MRST2007lomod PDFs.
The actual cross sections (without normalization rescaling factors) are shown.
\label{fig:W+}}
\end{figure}

\begin{figure}[ht]
\includegraphics[width=1.0\columnwidth,height=2\textheight,keepaspectratio]
{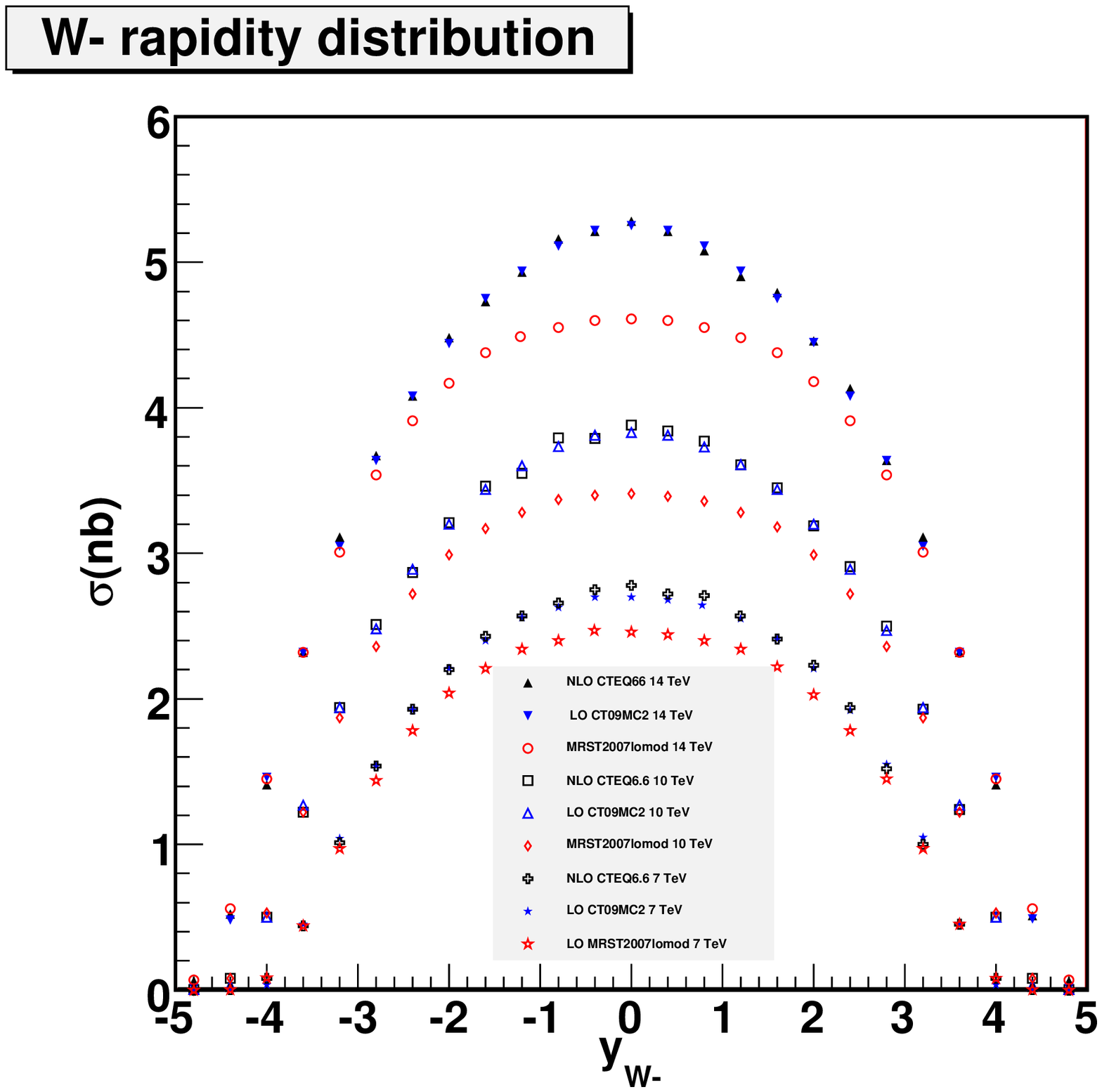}
\caption{Same as Fig.~\protect\ref{fig:W+}, for the $W^-$ rapidity distribution.
\label{fig:W-}}
\end{figure}

\begin{figure}[ht]
\includegraphics[width=1.0\columnwidth,height=2\textheight,keepaspectratio]
{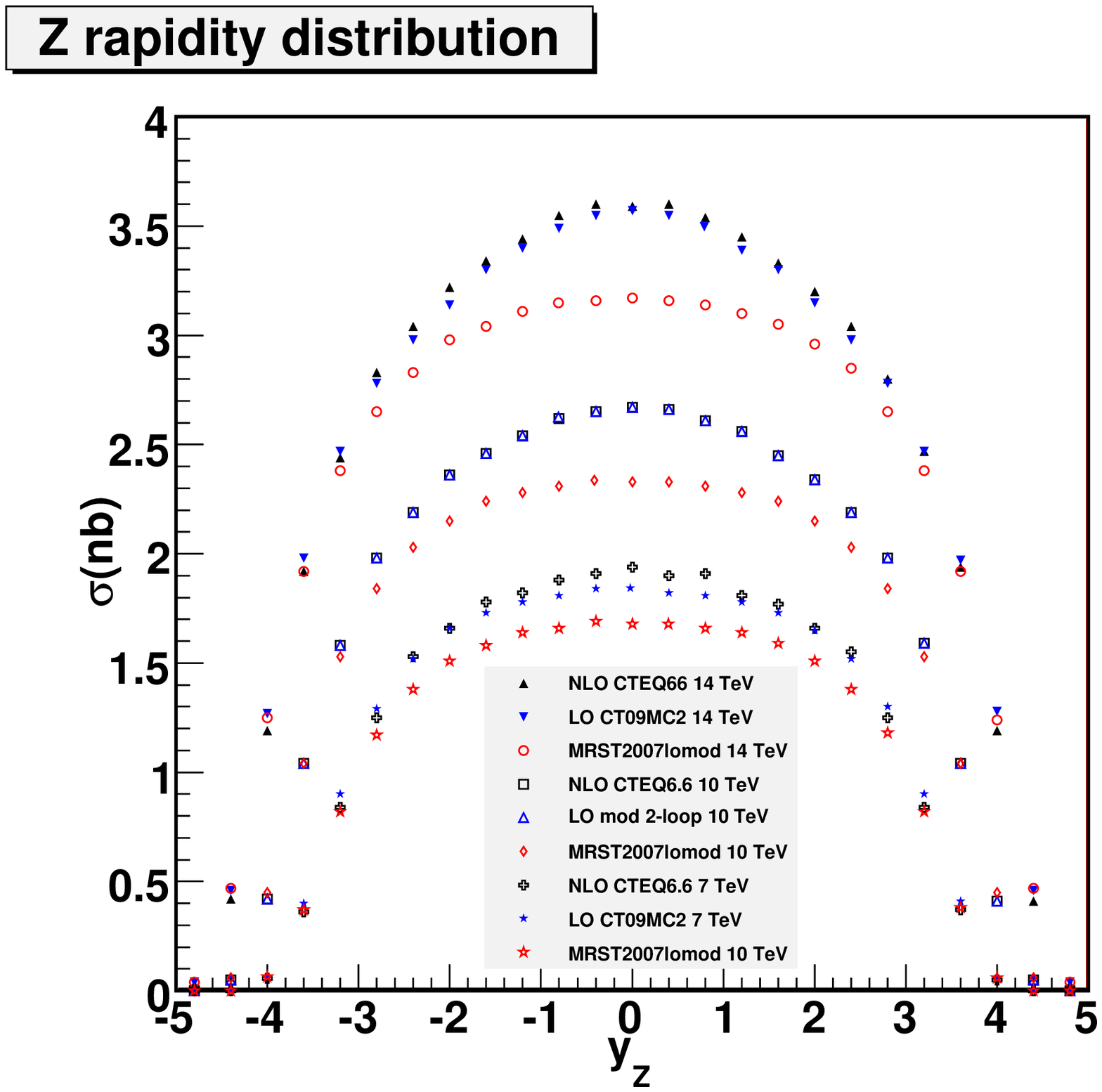}
\caption{Same as Fig.~\protect\ref{fig:W+}, for the $Z$ rapidity distribution.
\label{fig:Z}}
\end{figure}

\begin{figure}[ht]
\includegraphics[width=1.0\columnwidth,height=2\textheight,keepaspectratio]
{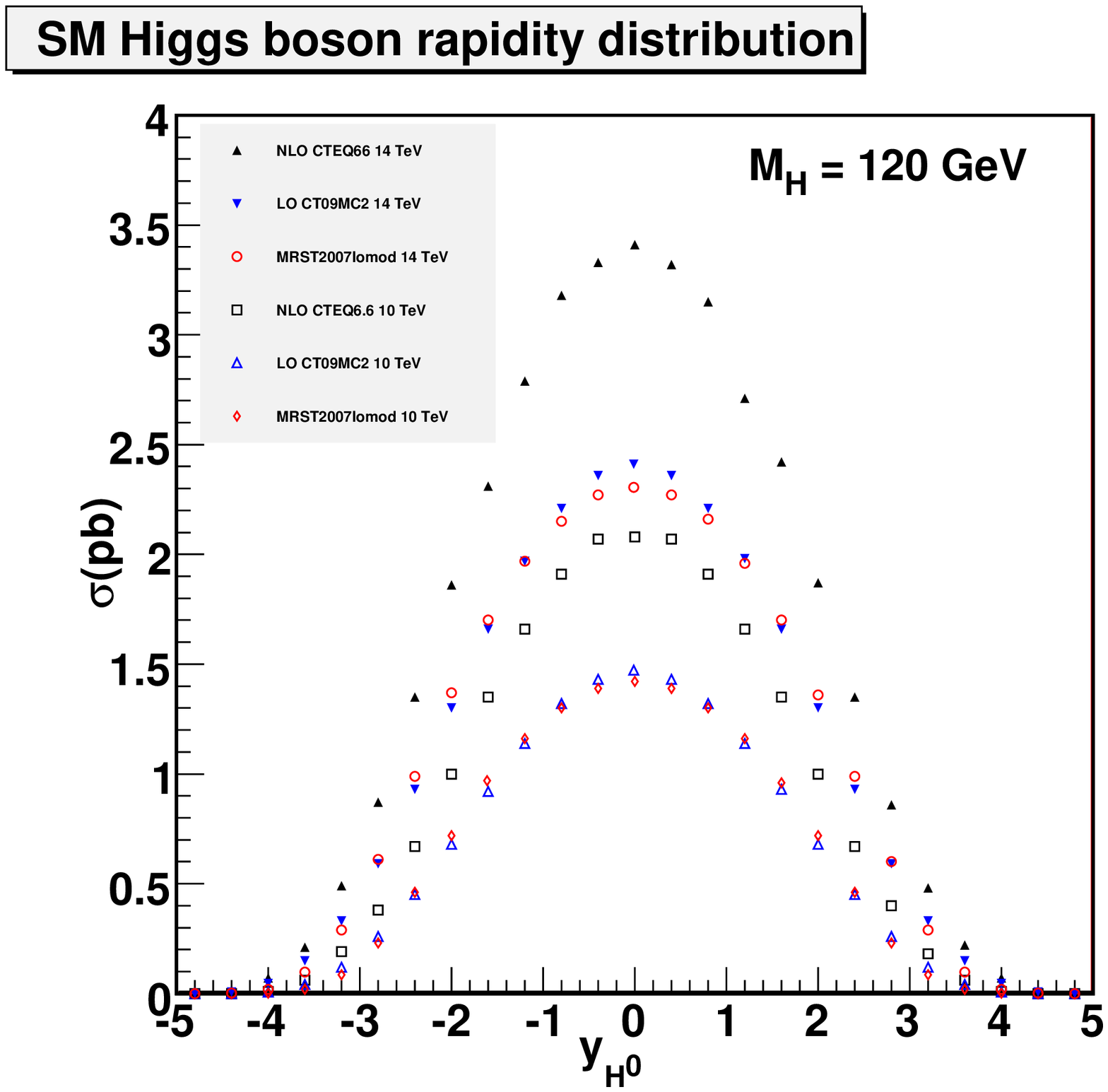}
\caption{Same as Fig.~\protect\ref{fig:W+}, for the Higgs boson rapidity distribution
at $\sqrt{s}=$10 and 14 TeV. 
To maintain legibility, the distribution for $\sqrt{s}=$7 TeV is not shown.
\label{fig:higgs}}
\end{figure}

\begin{figure}[ht]
\includegraphics[width=1.0\columnwidth,height=2\textheight,keepaspectratio]
{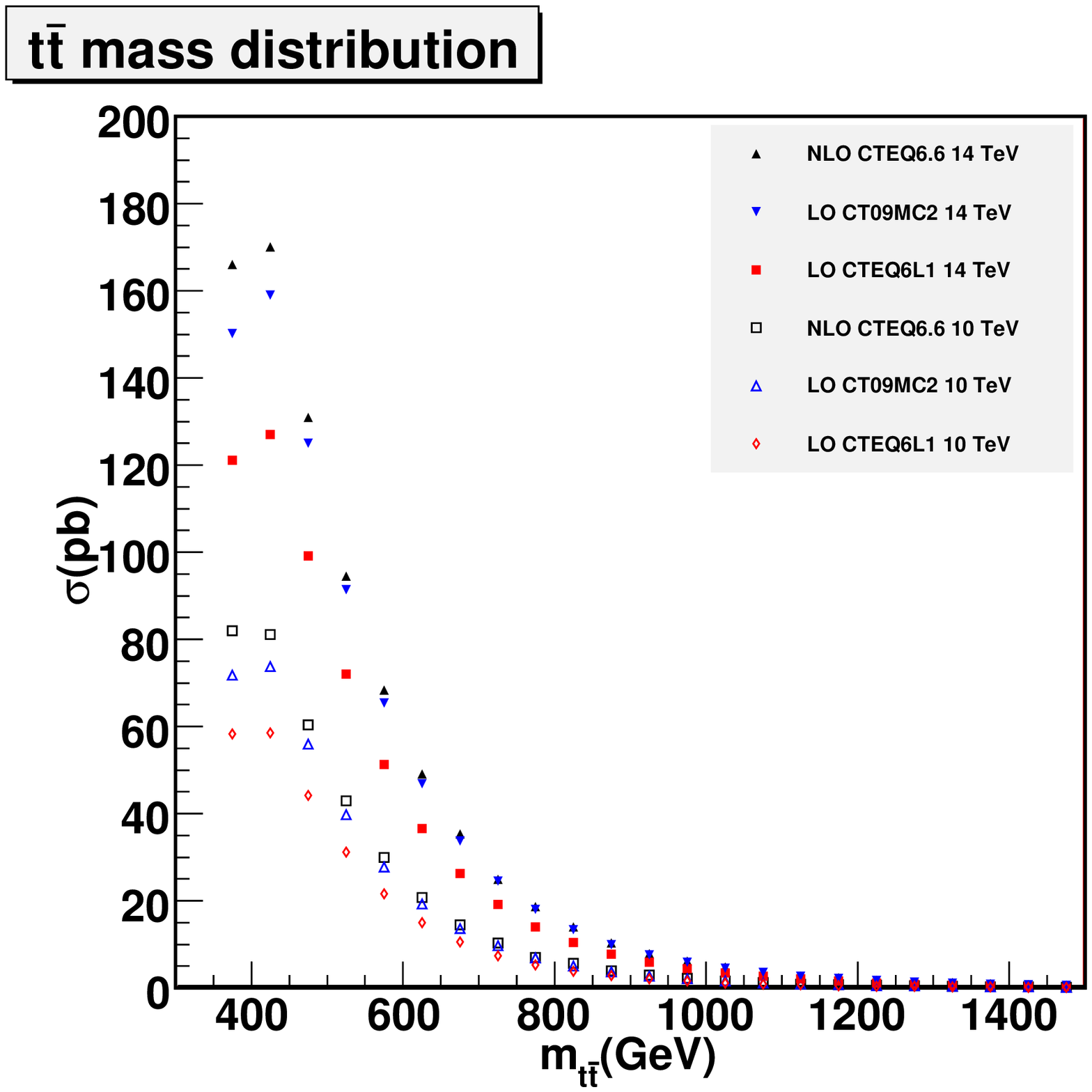}

\caption{
Predictions for the $t\bar{t}$ invariant mass distribution
at the LHC ($\sqrt{s}=$10 and 14 TeV) in 50 GeV mass bins,
given at NLO using the CTEQ6.6M PDFs,
and at LO using the CT09MC2 and CTEQ6L1 PDFs.
The actual cross sections are shown. 
\label{fig:tT}}
\end{figure}

\begin{figure}[ht]
\includegraphics[width=1.0\columnwidth,height=2\textheight,keepaspectratio]
{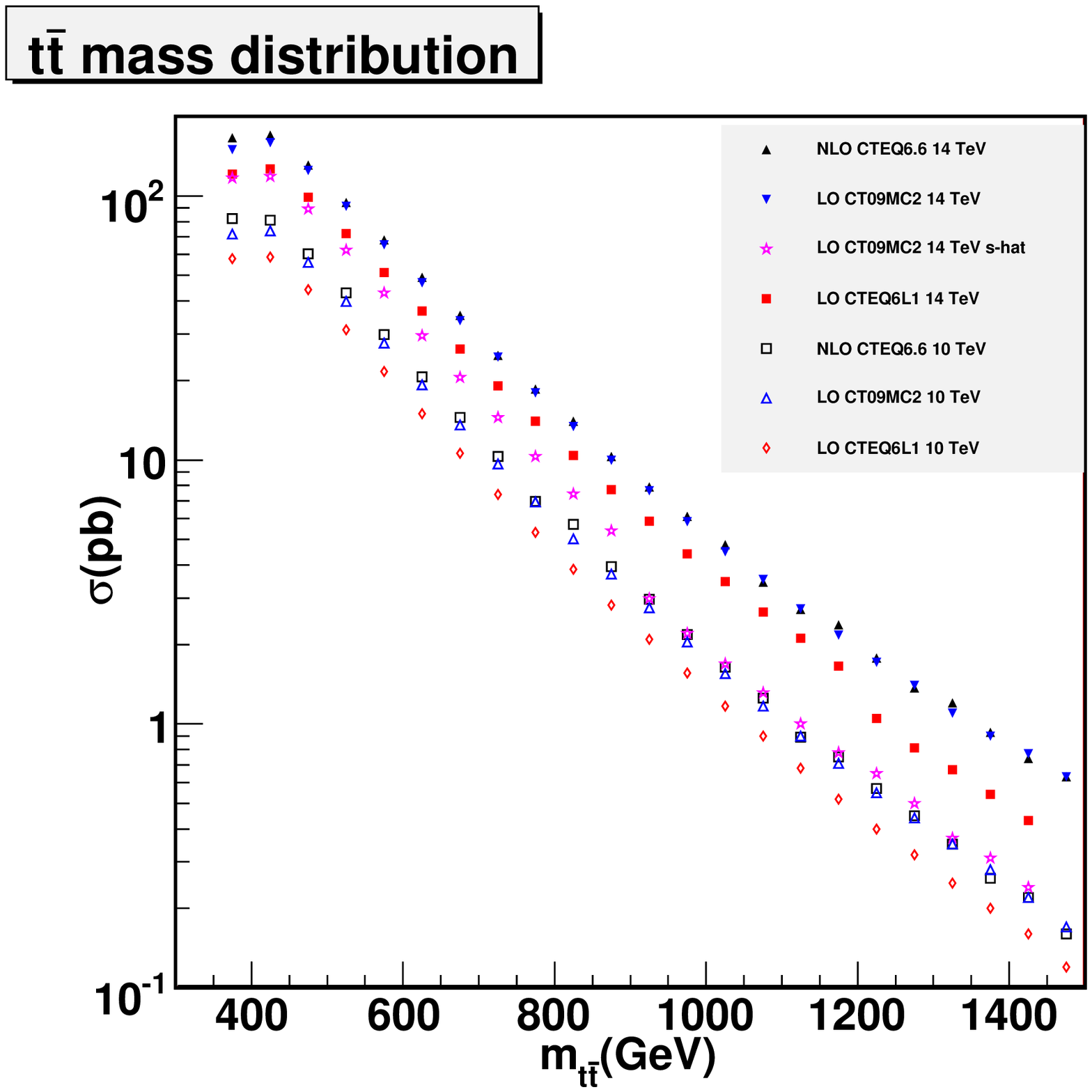}
\caption{
The same as in Fig.~\protect\ref{fig:tT}, on a semi-log scale. 
LO CT09MC2 predictions for the factorization scale 
$\mu=\sqrt{\widehat s}$ 
are also shown.
\label{fig:tT_log}
}
\end{figure}

Comparison of CT09MC2 predictions 
to the NLO pseudodata at the LHC center-of-mass energies $\sqrt{s}=$14, 10, 
and (for some processes) 7 TeV is shown in
Figs.~\ref{fig:W+}-\ref{fig:tT_log}. Similar values of cross
sections are obtained with the CT09MC1 PDF set. In the figures,
the actual cross sections are compared, without applying any
normalization factors.

In all cases, the LO cross sections based on the CT09MC2 PDFs 
are closer to the NLO predictions both
in the overall normalization and shape than the respective LO cross
sections based on a standard LO PDF such as CTEQ6L1.
The predictions for $W$ and $Z$ production at LO-MC
are almost identical to those at NLO,
and those for $t\bar{t}$ production are considerably closer
to the NLO predictions.\footnote{%
Both the LO and NLO predictions for $t\bar{t}$ production are evaluated
at the factorization scale $\mu=m_t$. 
The impact of using a different scale $\mu=\sqrt{\hat{s}}$ is also shown 
in Fig.~\protect\ref{fig:tT_log},
indicating the large scale dependence present in LO predictions.}
The predictions for the production of a 120 GeV Higgs boson 
are similar in shape, but the LO-MC prediction is still significantly lower 
than NLO (see the discussion below).
The LO-MC predictions have a similar or even better agreement
with the NLO benchmark cross sections at 
$\sqrt{s}=$7 and 10 TeV than at 14 TeV.

An alternative set of PDFs (MRST2007lomod) for leading-order Monte Carlo programs was
developed in Ref.~\cite{ThorneSh}. The figures compare the LO predictions utilising the  
MRST2007lomod PDFs with our results.
At 7 TeV, the difference between the LO predictions for $W$ and $Z$
production using the MRST2007lomod PDFs and the NLO benchmark cross sections
is essentially a normalization shift. At 10 TeV, and then especially at 14 TeV,
there is also a noticeable difference in the shape of the rapidity distribution.
However, both CT09MC2 and MRST2007lomod predictions 
provide an almost identical description for Higgs production. 

The K-factors that need to be applied to the LO CT09MC1 and CT09MC2 predictions 
to recoincile them with their NLO  counterparts are listed in Table 2. 
The K-factors for $W$ and $Z$ boson production are basically unity,
made possible by the extra freedom introduced
by the relaxation of the momentum sum rule.
The K-factors for the gluon-induced processes are closer to unity than for
the standard LO PDF,
as a result of the larger gluon density at high $x$.
The K-factor for Higgs production remains significantly larger than unity,
since the virtual corrections to this process 
are especially large and cannot (nor should they)  be
completely compensated by an increase in the LO gluon density. 

\clearpage

\begin{table}[ht]
\begin{center}
\begin{tabular}{|c|c|c|c|c|c|c|c|}
\hline 
  & $W^{+}$  & $W^{-}$  & $Z$  & $H$  & $t\bar{t}$
  & $b'\overline{b'}$ & momentum sum
\tabularnewline
\hline
\hline 
$K_{i}$ (CT09MC1)  & 1.00  & 0.99  & 0.98  & 1.22  & 1.09  & 2.70 & 1.10
\tabularnewline
\hline
$K_{i}$ (CT09MC2)  & 1.02  & 1.00  & 1.00  & 1.32  & 1.09  & 3.13 & 1.14 
\tabularnewline
\hline
\end{tabular}
\caption{Fitted $K_{i}$ for each pseudodata set at the LHC (at 14 TeV)
 for CT09MC1 and CT09MC2 PDFs,
along with the sum of parton momentum fractions in the proton for each set.
\label{tab:fitted-LHC-norm}}
\end{center}
\end{table}

\section{Comparisons of PDFs}

\begin{figure}
\begin{center}
\includegraphics[width=7cm,angle=0]{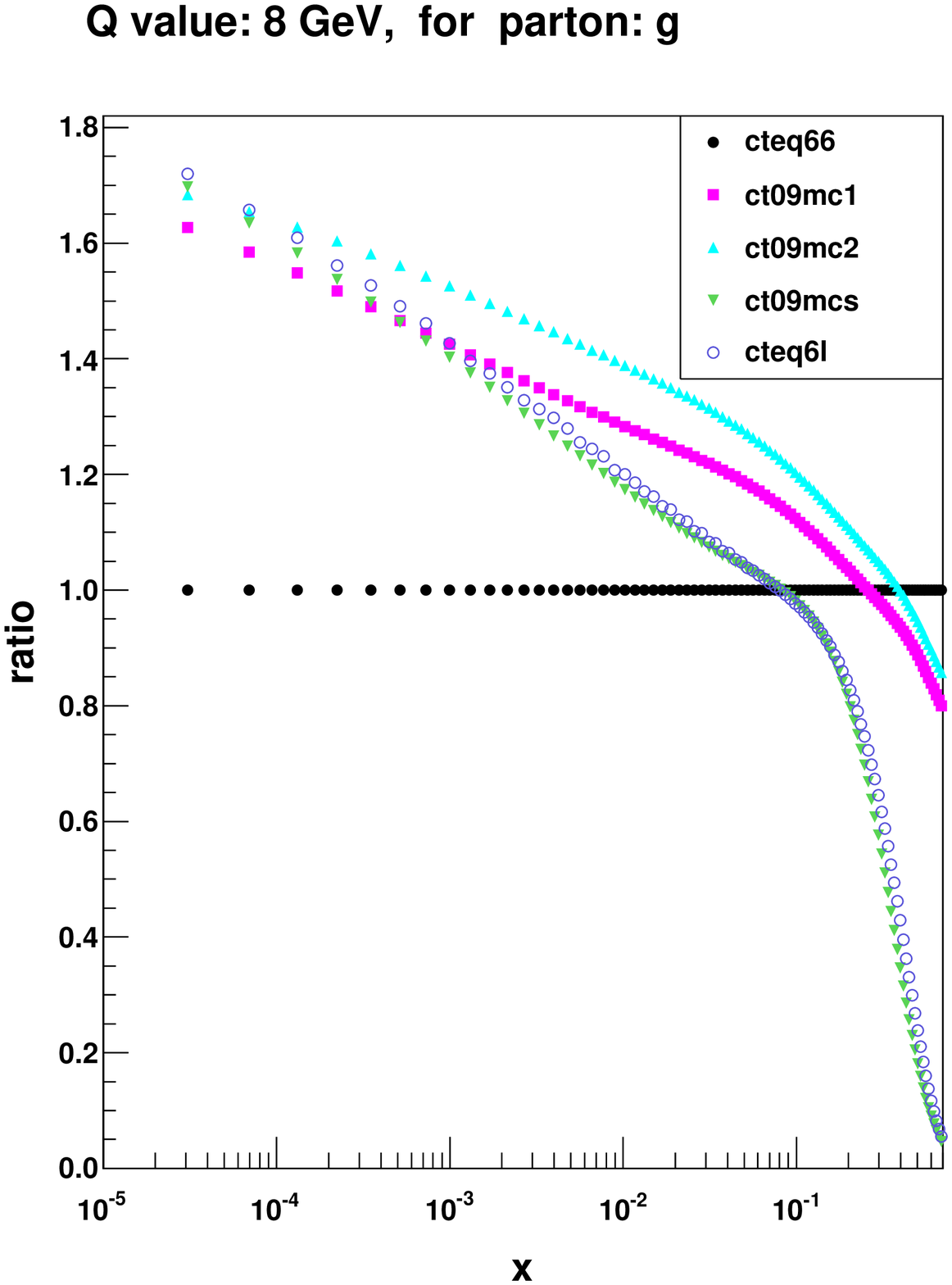}
\includegraphics[width=7cm,angle=0]{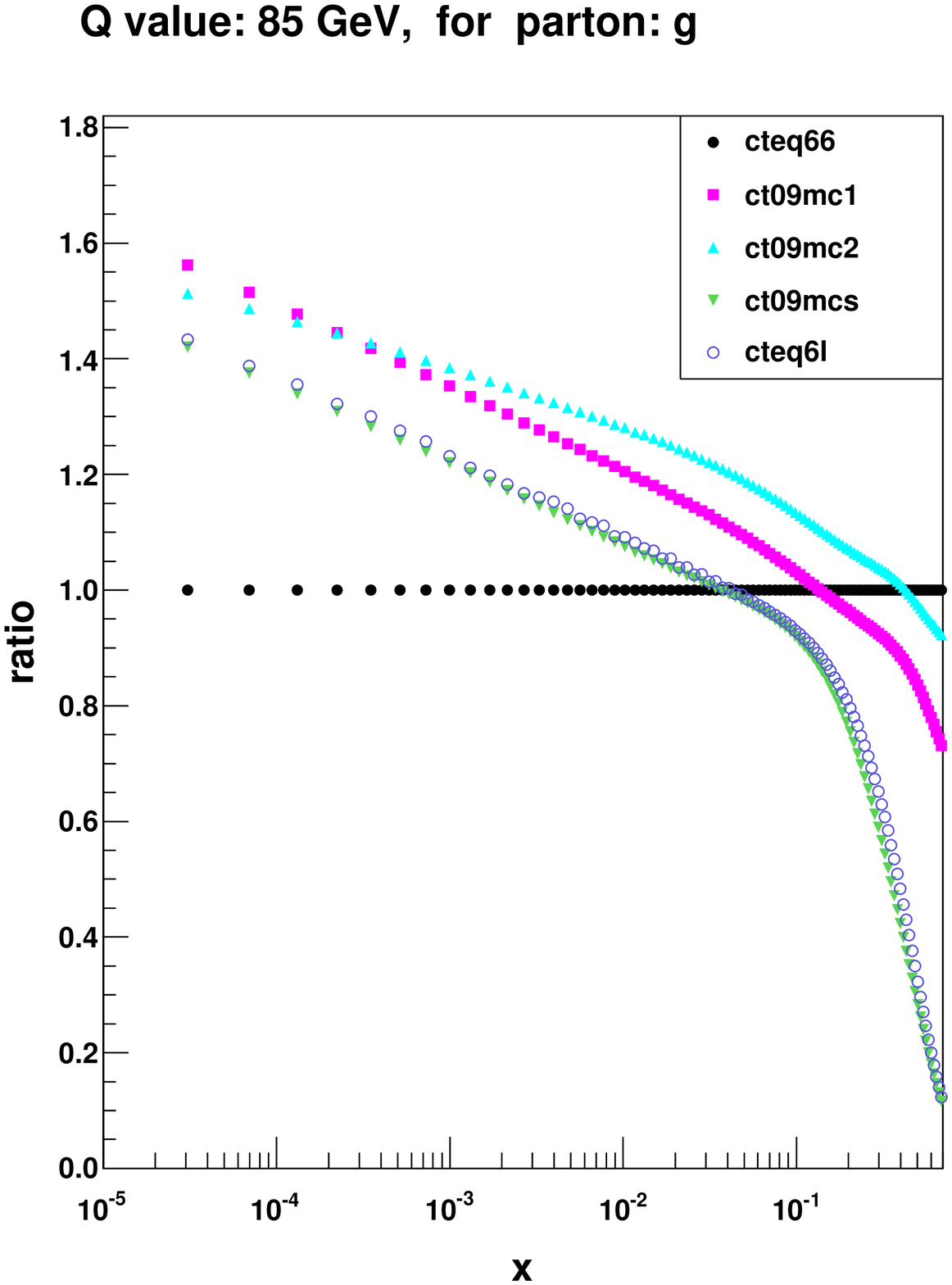}
\end{center}
\vspace*{-0.5cm}
\caption{The ratio of gluon distributions from various LO PDFs
to the gluon distribution from CTEQ6.6M at $Q$ values of 8 and 85 GeV. 
\label{fig:gluon_28}}
\end{figure}

\begin{figure}
\begin{center}
\includegraphics[width=7cm,angle=0]{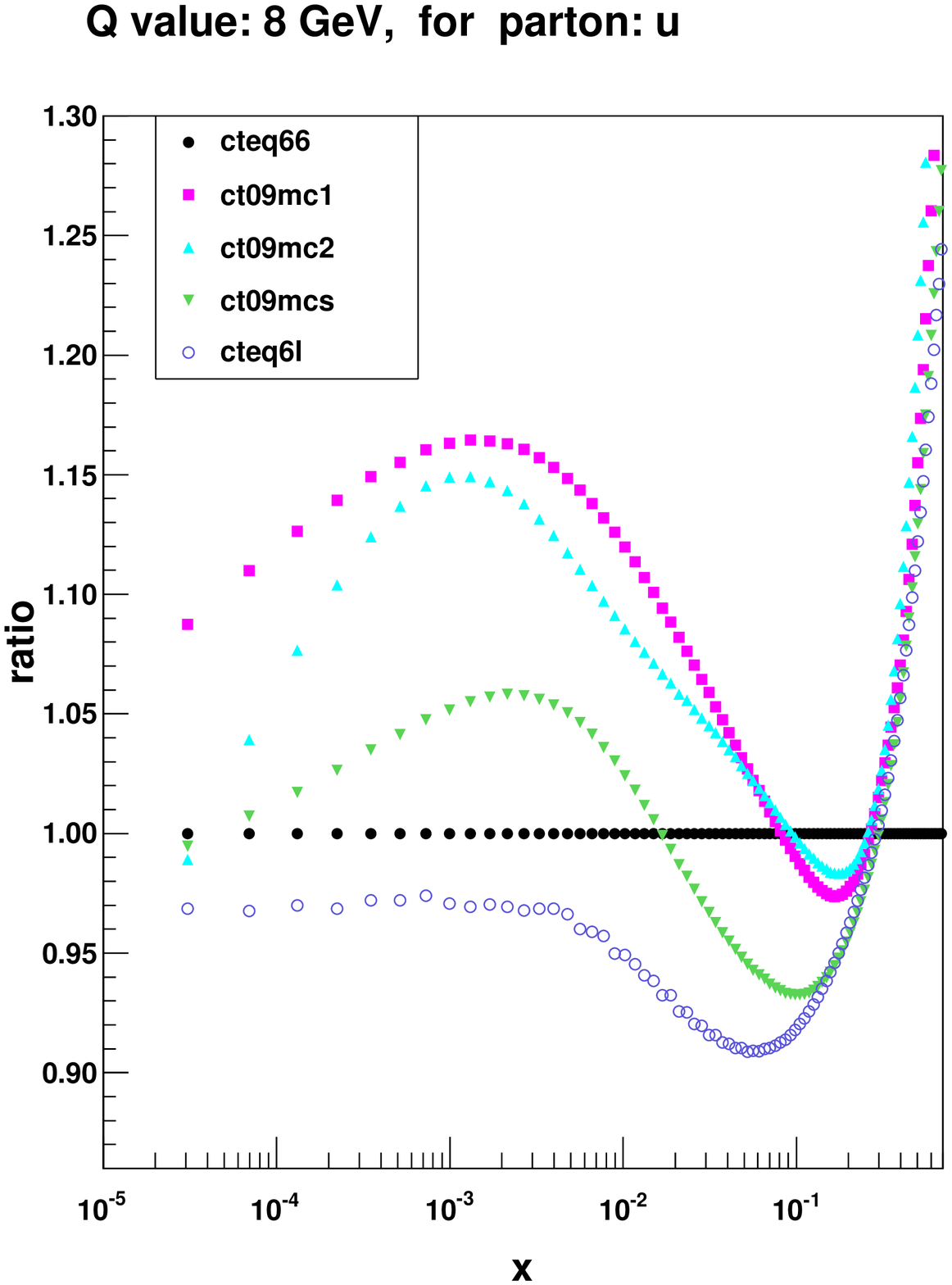}
\includegraphics[width=7cm,angle=0]{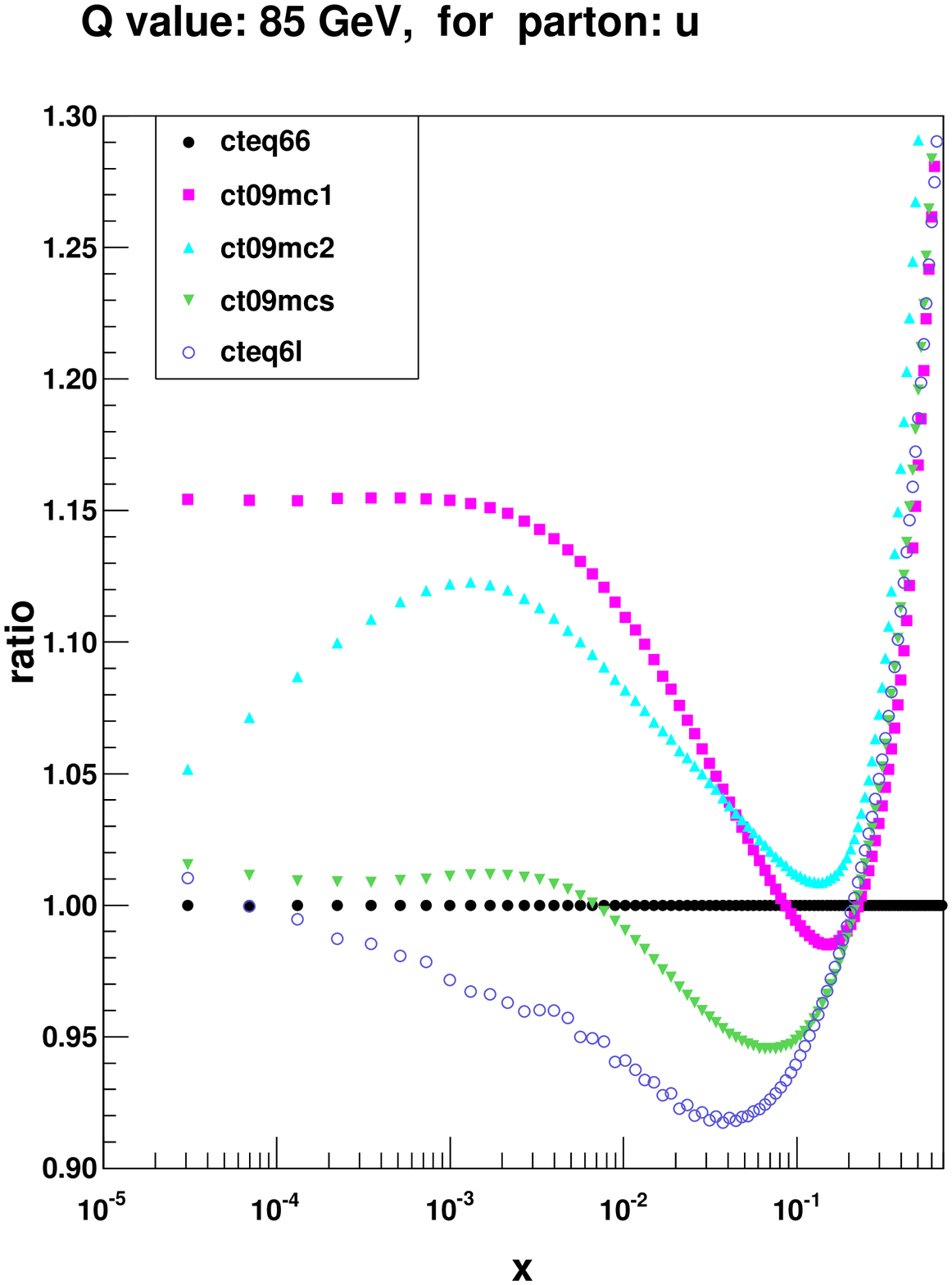}
\end{center}
\vspace*{-0.5cm}
\caption{The ratio of the $u$ quark distributions from various LO PDFs
to the $u$ quark  distribution from CTEQ6.6M at $Q$ values of 8 and 85 GeV. 
\label{fig:up_8}}
\end{figure}

\begin{figure}
\begin{center}
\includegraphics[width=7cm,angle=0]{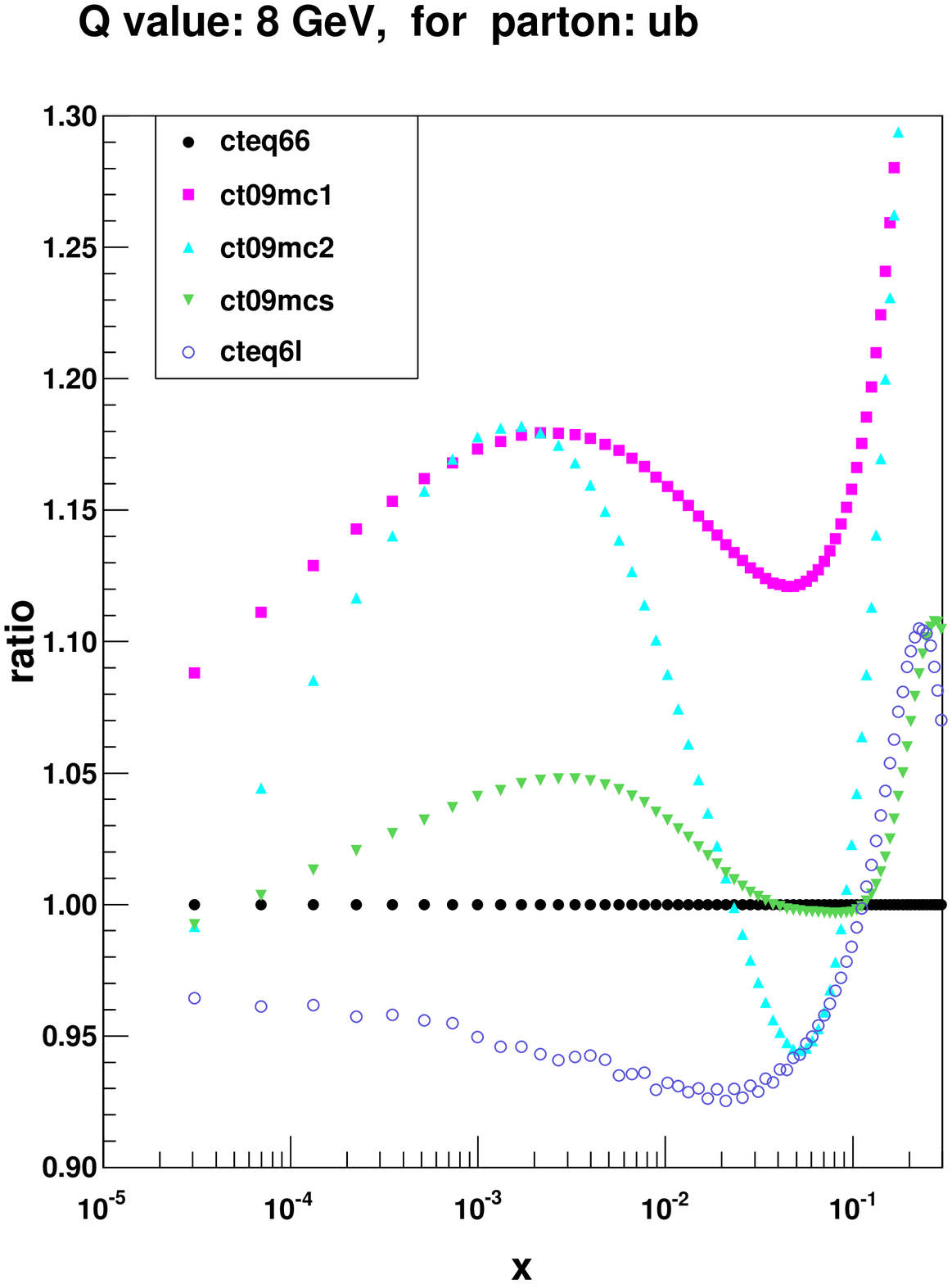}
\includegraphics[width=7cm,angle=0]{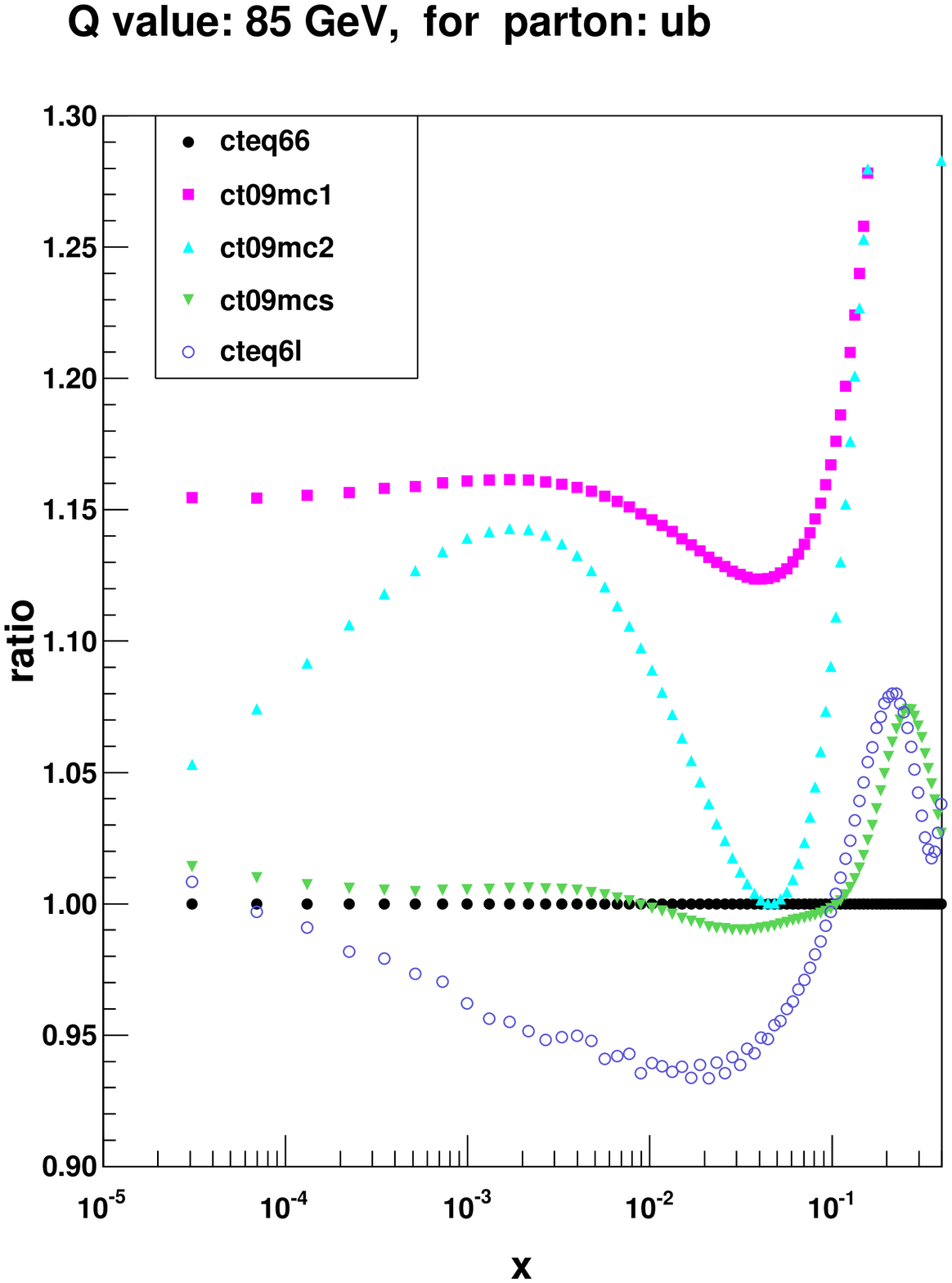}
\end{center}
\vspace*{-0.5cm}
\caption{The ratio of the $\bar{u}$ distributions from various LO PDFs
to the $\bar{u}$ distribution from CTEQ6.6M at $Q$ values of 8 and 85 GeV. 
\label{fig:ubar_8}}
\end{figure}

\begin{figure}
\begin{center}
\includegraphics[width=7cm,angle=0]{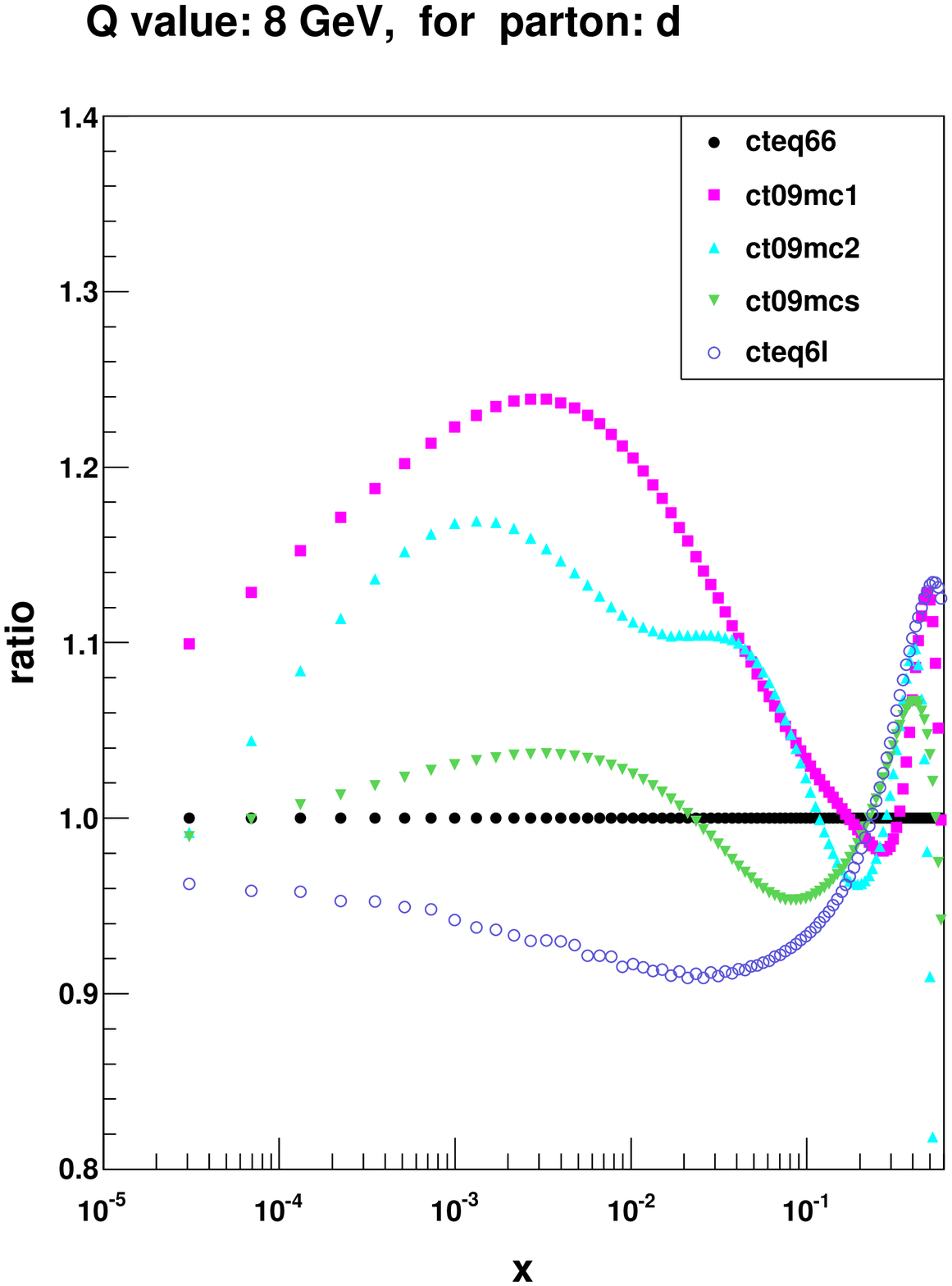}
\includegraphics[width=7cm,angle=0]{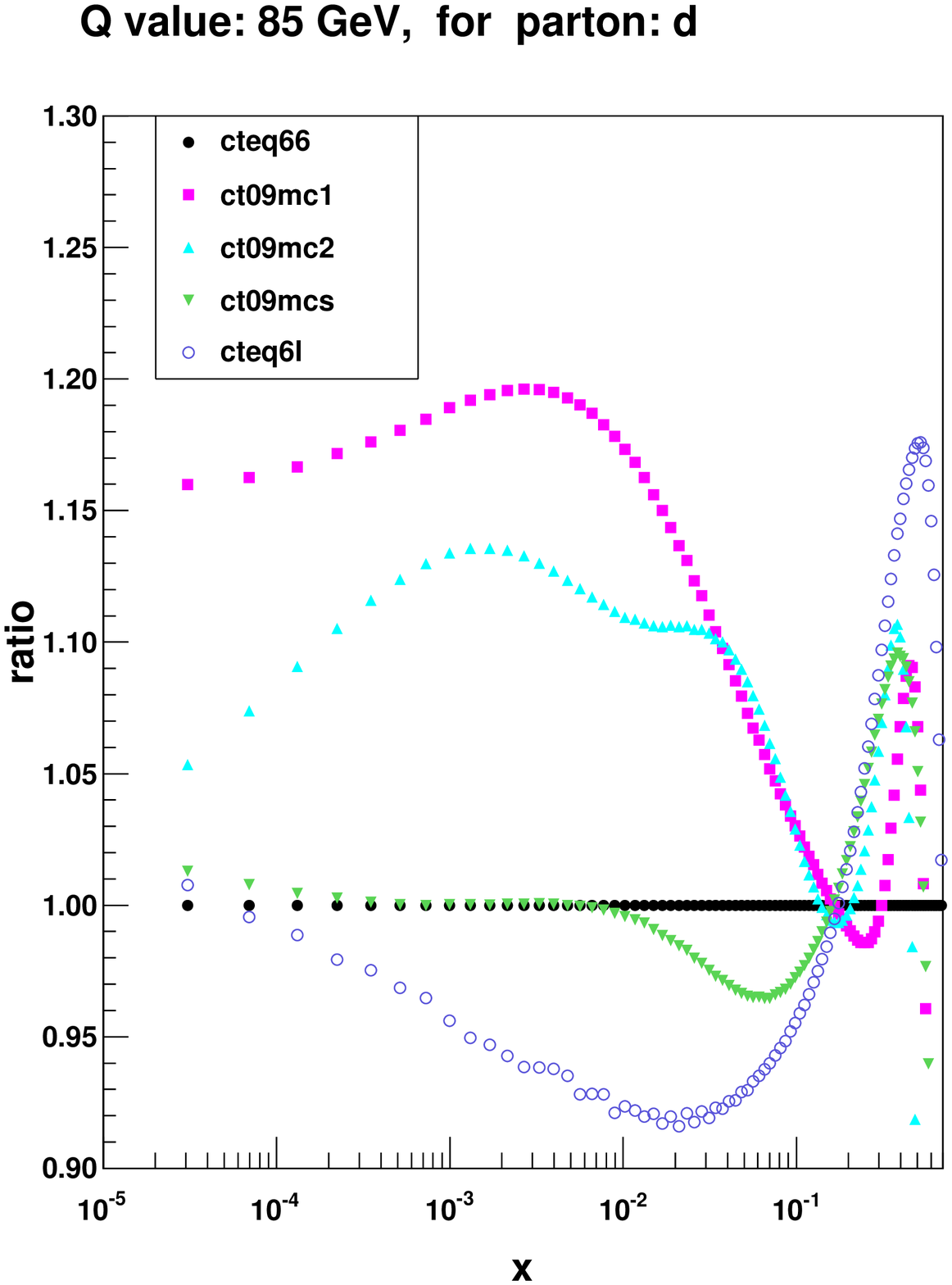}
\end{center}
\vspace*{-0.5cm}
\caption{The ratio of the $d$ quark distributions from various LO PDFs
to the $d$ quark distribution from CTEQ6.6M at $Q$ values of 8 and 85 GeV. 
\label{fig:down_8}}
\end{figure}

Figures~\ref{fig:gluon_28}-\ref{fig:down_8} compare 
the LO-MC PDFs described in this paper with the CTEQ6.6M and CTEQ6L PDFs,
for various parton flavors and energy scales. The LO-MC gluon PDF CT09MCS,
obtained with the fitted normalizations and scales,
 is quite close to the conventional LO PDF, CTEQ6L, 
as seen in Fig.~\ref{fig:gluon_28}. The gluon distributions 
in two LO-MC fits with the relaxed momentum sum rule, 
CT09MC1 and CT09MC2, are equal to, or larger than, CTEQ6L in the entire
 $x$ range. They are larger than the CTEQ6.6M gluon up to $x$ values 
of 0.1 (CT09MC1) and 0.4 (CT09MC2).
All LO-MC gluon PDFs approach the CTEQ6L gluon PDF at small $x$
($0.001$ or less), in the region responsible for producing 
the underlying event at the LHC. 
With the momentum sum rule relaxed, the 2-loop CT09MC2 gluon is
 noticeably larger than the 1-loop CT09MC1 gluon, in order to compensate
 for the smaller value of the 2-loop QCD coupling strength when
 fitting the NLO pseudodata.\footnote{Such increase does not 
 happen if the momentum sum rule is enforced. For example, the
CTEQ6L1 gluon PDF (with 1-loop $\alpha_s$) is about the same 
as the CTEQ6L gluon PDF (with 2-loop $\alpha_s$).}

The increase in the CT09MC1 and CT09MC2 gluon distributions 
is accompanied by the significant increase in the small-$x$ 
sea quark distributions. The LO-MC $u$-quark distributions 
(cf. Figure \ref{fig:up_8}) 
remain larger than the NLO $u$-quark distribution at $x>0.2$,
in a manner similar to the conventional CTEQ6L $u$-quark distribution.
The $u$,  $\bar{u}$, ${d}$, and $\bar{d}$
distributions are larger than CTEQ6.6M and CTEQ6L at small and moderate $x$,
leading to both a flattening of the LHC 
$W^+$ rapidity distribution and an increase in the total cross
sections for the vector boson pseudodata required by the 
full NLO calculations. Finally, while the CT09MC PDFs for
(anti)quarks are quite different from their MRST2007lomod
counterparts, the CT09MC gluon distributions 
are similar to those from MRST except at high $x$,
where the CT09MC PDFs are closer to CTEQ6.6M
due to the influence of the pseudodata. 

\begin{figure}
\begin{center}
\includegraphics[width=7cm,angle=0]{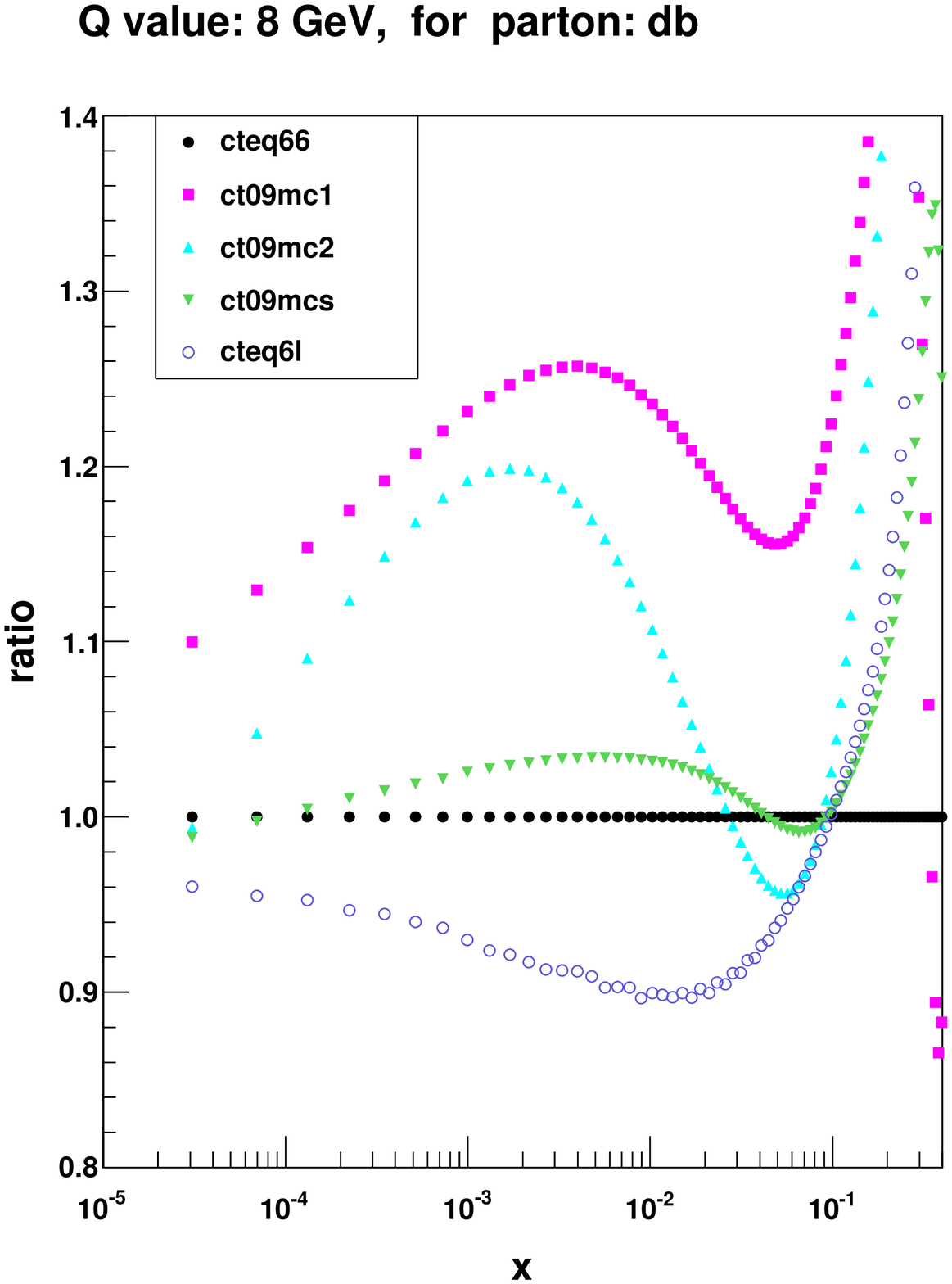}
\includegraphics[width=7cm,angle=0]{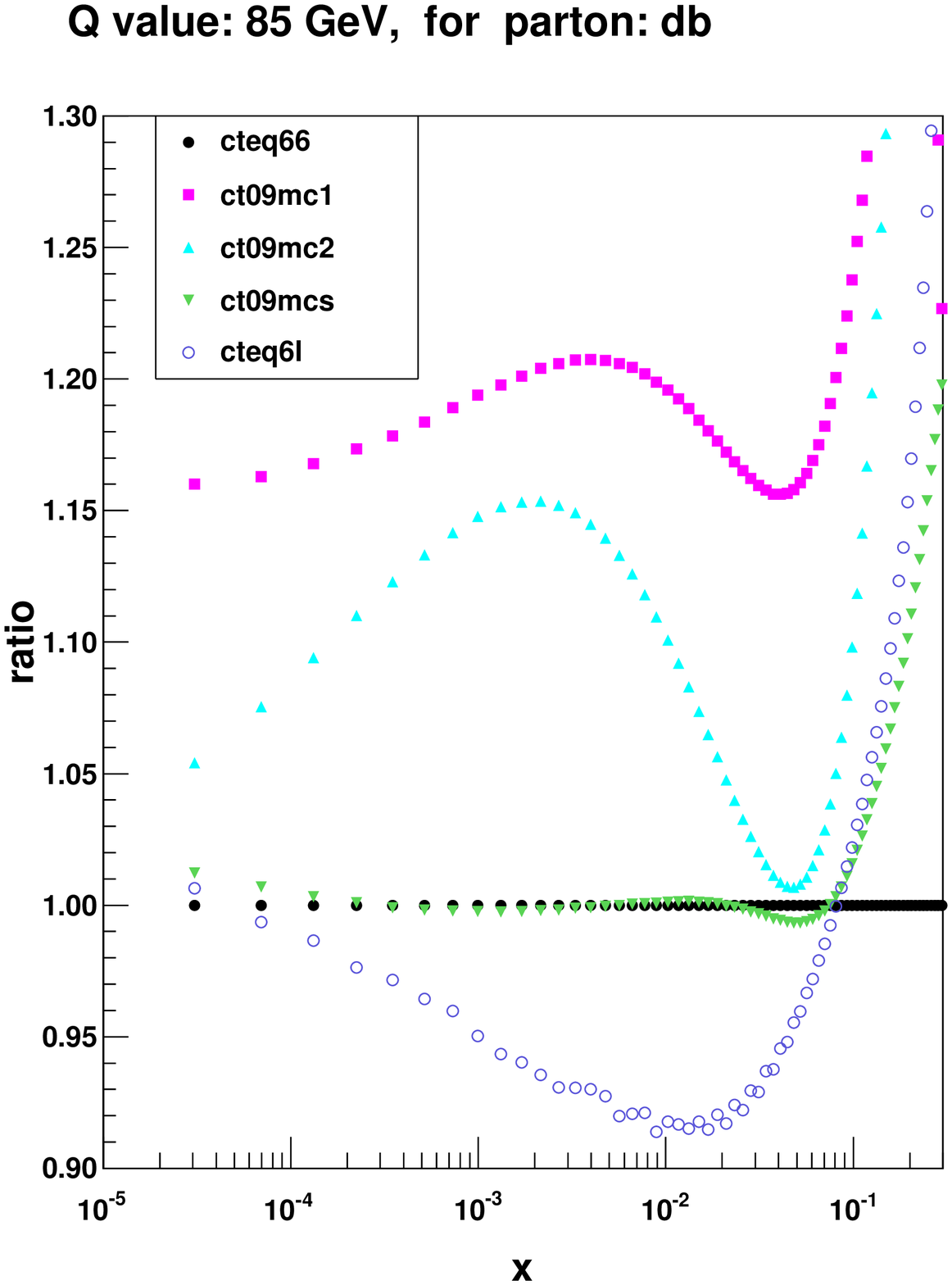}
\end{center}
\vspace*{-0.5cm}
\caption{The ratio of the $\bar{d}$ distributions from various LO PDFs
to the $\bar{d}$ distribution from CTEQ6.6M at $Q$ values of 8 and 85 GeV. 
\label{fig:dbar_8}}
\end{figure}

\clearpage

\section{Predictions for other LHC cross sections
\label{sec:OtherLHCXsec}}

\begin{figure}
\begin{centering}
\includegraphics[width=0.8\columnwidth,height=1.6\textheight,keepaspectratio]
{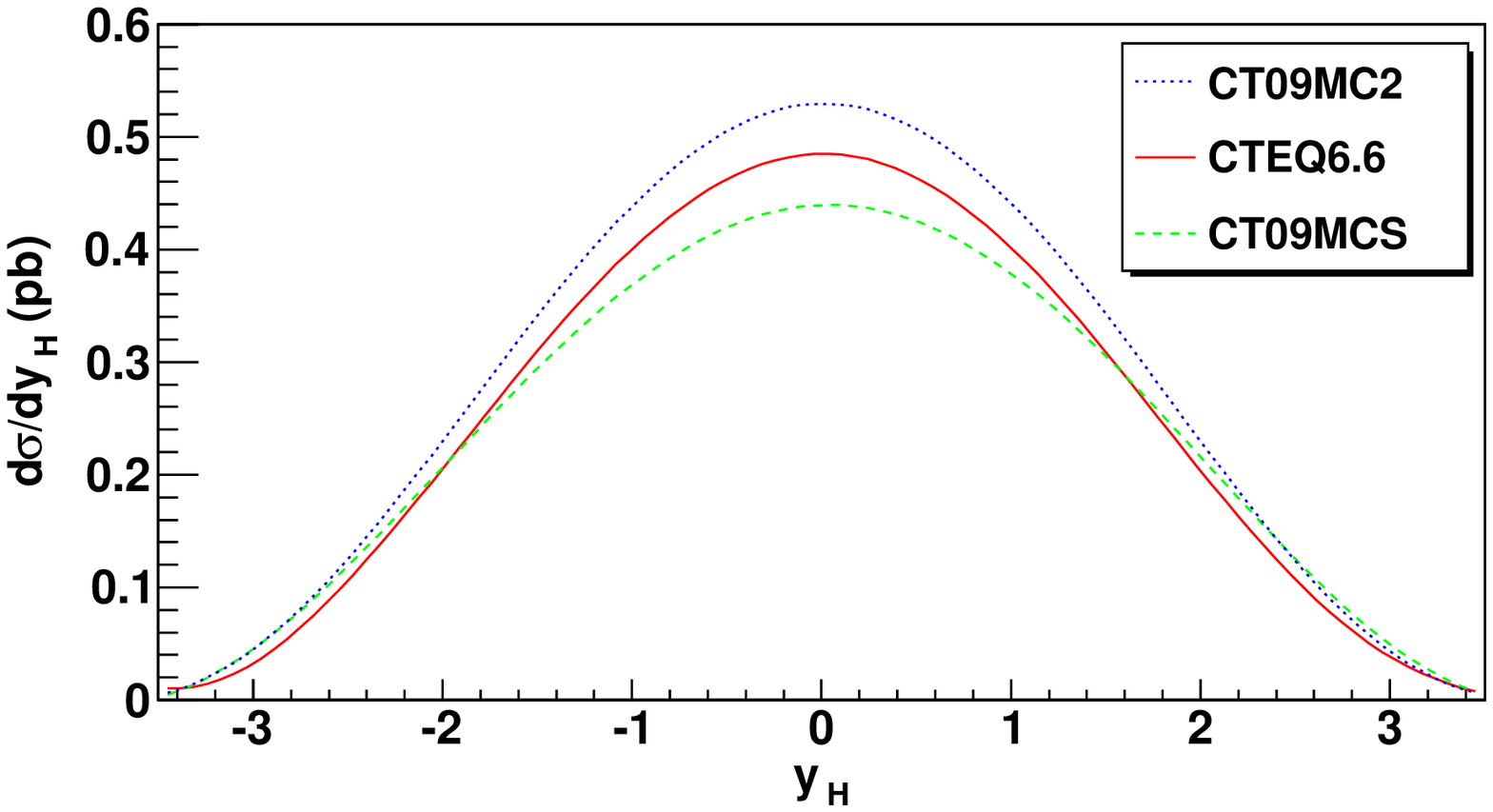}
\includegraphics[width=0.8\columnwidth,height=1.6\textheight,keepaspectratio]
{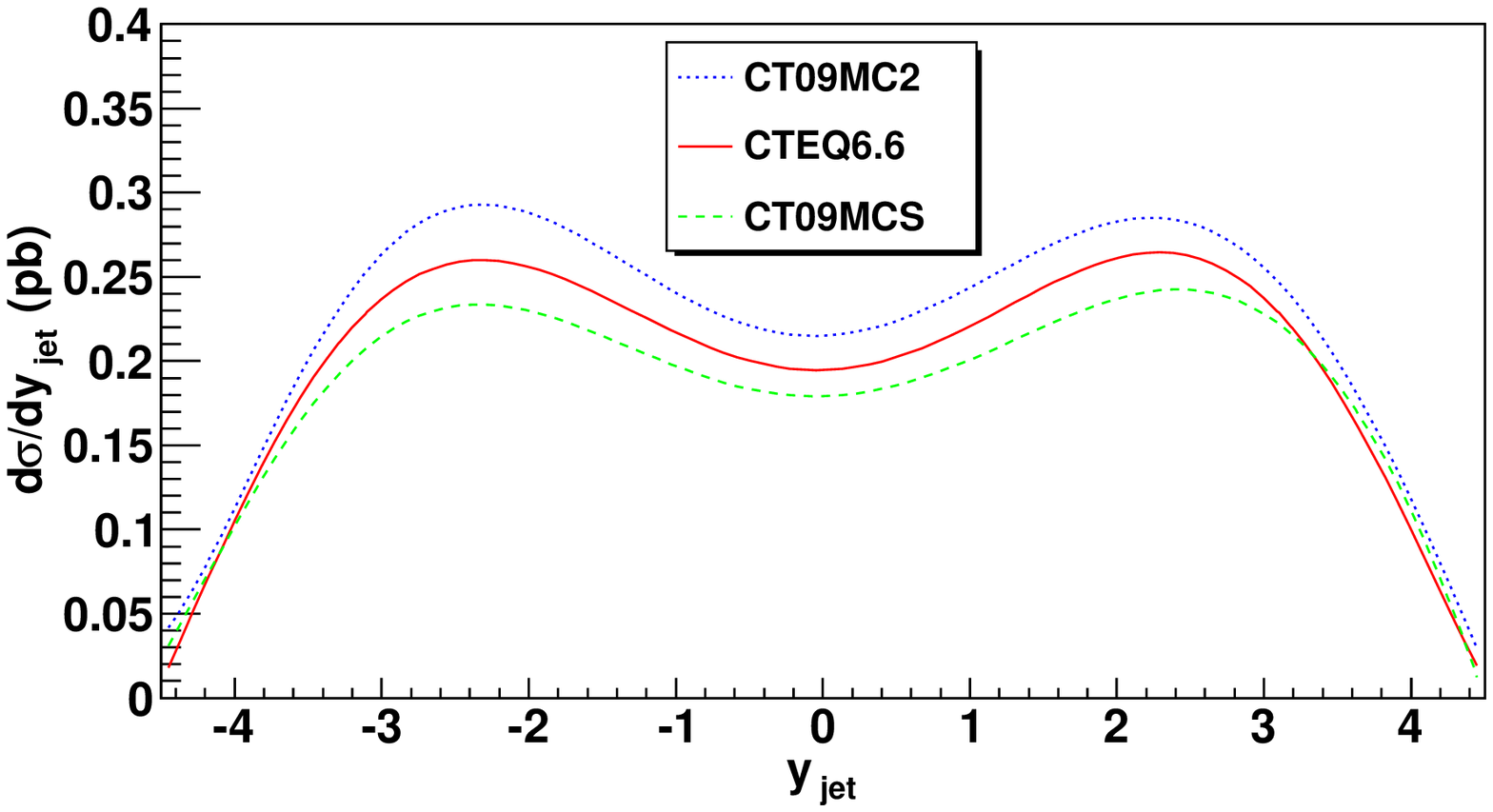}
\caption{The rapidity distribution of 120 GeV Higgs bosons produced
through vector boson fusion at $\sqrt{s}=$14 TeV (top).
Also shown is the distribution in the rapidity of the leading jet (bottom). 
NLO predictions are obtained with the CTEQ6.6M PDFs (solid curves), 
and LO predictions are for the CT09MCS (dashed curves) 
and CT09MC2 (dotted curves) PDFs. Here, 
the jets are separated by $\Delta R>0.4$ 
(with $R_{sep}=1.3$), and  the transverse momentum and pseudorapidity of the
jet satisfy $p_T > 40\ \mathrm{GeV}/c$
and $|\eta|<4.5$.
\label{fig:vbf}}
\end{centering}
\end{figure}

By construction, predictions based on the LO-MC PDFs
provide a better description of the LHC pseudodata cross sections.
The pseudodata sets were chosen so as to be representative of 
the universally desired PDF behavior for typical
LHC hard-scattering cross sections, 
but it is important to check predictions for the cross sections
that were not a part of the pseudodata sets.
In Fig.~\ref{fig:vbf},
we show cross sections for vector boson fusion production of a SM Higgs boson
($m_{\mathrm{Higgs}}=120\,{\mathrm{GeV}}$), computed
at NLO using the CTEQ6.6M PDFs, 
and at LO using two LO-MC PDFs (CT09MCS and CT09MC2). Distributions in 
the rapidities of the Higgs
boson and the leading jet are plotted. The two LO-MC calculations reproduce well
the shapes of the NLO rapidity distributions.
The LO-MC cross sections are larger (smaller) 
than the respective NLO cross sections when the CT09MC2 (CT09MCS) PDFs are used.
Both of them differ from the NLO (CTEQ6.6M) 
prediction in the central rapidity region by about ten percent.

To study the impact of the LO-MC
PDFs on the matching of (multi-parton) hard matrix elements with parton showers, 
we have performed a comparison of parton-level cross sections 
for production of $W^+$ + n-partons (n=0,..,4) at the LHC (10 TeV), computed by 
the MADGRAPH program~\cite{Alwall:2007st} with both the conventional (CTEQ6L1) and
CT09MC2 PDFs. We have found that the CT09MC2 PDFs increase the
subprocess cross sections by a factor of about 1.25-1.35 for
 $q\bar{q}$ and $gq$ initial states, 
relatively independently of the flavors of the initial-state quarks 
and the number of partons in the final state. 
For $gg$ initial states, the factor is larger, ranging from about 1.5 to 1.75. 
The details of this comparison are collected in Appendix A.

The K-factor for a given process is a useful shorthand which encapsulates
the size of the NLO corrections to the lowest-order cross section.
As discussed in Appendix B, the K-factors are closer to unity when NLO PDFs 
are used for the LO calculations, and this is true as well for LO predictions 
using the CT09MC1 and CT09MC2 PDFs.

\section{Impact on the underlying event at the LHC}

\begin{figure}[ht]
\begin{centering}
\includegraphics[width=0.7\columnwidth,height=1.4\textheight,keepaspectratio]
{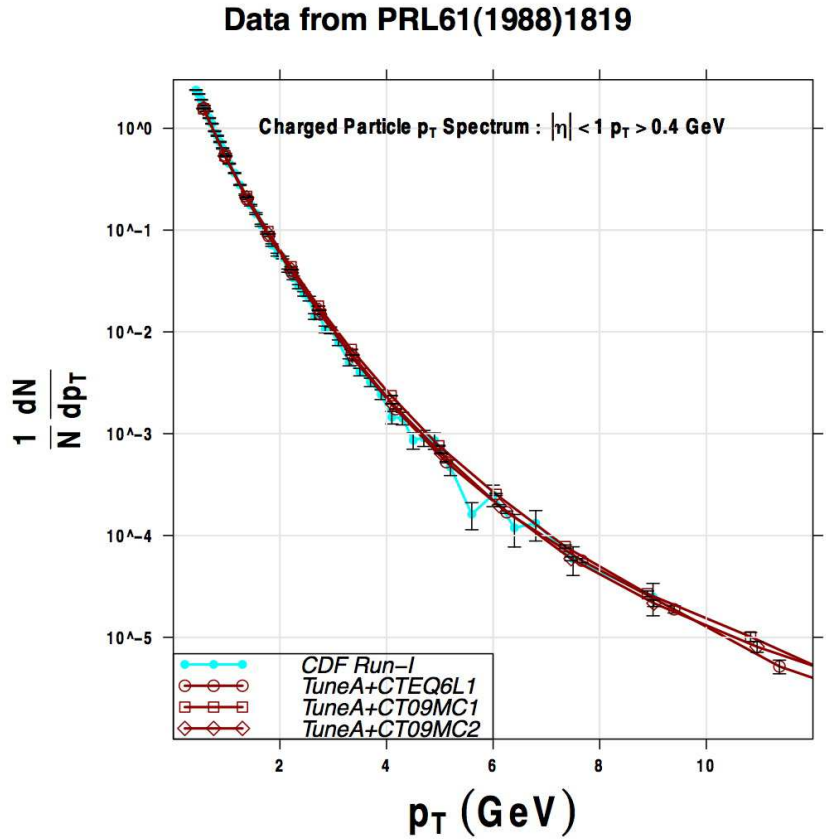}
\caption{Predictions for the charged particle transverse momentum distribution in minimum bias events for CDF in the Tevatron Run 1 (1.8 TeV), using the CTEQ6L1, CT09MC1 and CT09MC2 PDFs. 
 \label{fig:cdf}}

\end{centering}
\end{figure}

Predictions for the underlying event at the LHC are most sensitive
to the magnitude and shape of the low-$x$ gluon PDF,
as the small-$x$ $gg$ scattering into low-$p_T$
dijets makes up the bulk of the underlying event.
As stated earlier,
the LO gluon distribution is considerably larger at low $x$ than the NLO gluon.
The multiple parton scattering models in the LO parton shower Monte Carlos
have been tuned to this default LO gluon behavior.
The CT09MC PDFs retain the low-$x$ behavior of the conventional LO gluon PDF
and thus can be used with underlying event tunes similar to those derived
for standard LO PDFs~\cite{markus}. As an example, Fig.~\ref{fig:cdf}
 shows PYTHIA~\cite{Sjostrand:2006za}
 predictions for the charged particle transverse momentum distribution
 in minimum bias events at CDF, obtained for PYTHIA Tune A \cite{tuneA} 
 and CTEQ6L1, CT09MC1, and CT09MC2 PDFs. The two L0-MC PDFs lead to an equivalent 
 description of the standard, Tune A with CTEQ6L1. 

\section{Conclusion\label{sec:summ}}

In this paper, we have generalized the conventional global QCD analysis to
produce parton distributions optimized for simulations in event generators at
leading order in perturbative QCD.
This is done by combining the constraints due to existing hard-scattering
experimental data with those from anticipated cross sections for
key representative SM processes at the LHC 
(predicted by the NLO QCD theory) as a joint input to the global analysis.
Results obtained from a few candidate PDF sets for LO event generators
produced this way have been compared with those from other approaches.
As compared to the conventional LO PDFs, 
the PDFs for leading-order Monte-Carlo event generators (LO-MC PDFs) 
described here provide a better description
for the normalization of the benchmark LHC cross sections,
but, more importantly, for the shapes of these cross sections. 
In addition, we have performed validation studies to gauge the
phenomenological impact of the CT09MC PDF sets and to locate any possible pathological
behavior. Aside from the (desired) differences with the conventional
LO PDFs noted in this paper, the effects are otherwise benign. In
particular, the CT09MC PDF sets can be used with the underlying event
tunes similar to those performed with CTEQ6L1. For the LHC 
processes discussed
in Section~\ref{sec:OtherLHCXsec}, we have checked 
kinematic properties of parton-level jets obtained with
the new PDFs. After considering the individual $p_T$ and rapidity values of the
jets, as well as variables sensitive to correlations between the jets, 
such as $m_{jj}$, $\Delta R(j,j)$, etc., no unexpected features were observed
beyond the usual differences due to the choice of different PDF sets. 

Given their good agreement with the anticipated LHC cross sections,
the resulting PDFs are intended primarily for simulations for the LHC,
and only using LO event generators. This study is our first attempt to develop
such optimal PDFs. The discussed approach, and the choices made,
are only representative of what can be achieved with this method.
Given the very nature of the LO event
generators themselves, and the inherent uncertainties of any calculation
done with LO matrix elements, it is the distinctive qualitative
features of these PDFs described in earlier sections that
matter the most.

\section{In Memoriam}
This paper would not have been possible without the insight and inspiration
of our late colleague Wu-Ki Tung. 

\paragraph{Acknowledgements}
We would like to thank Albert de Roeck, Hannes Jung, Judith Katzy, Peter Skands,
Torbjorn Sjostrand, Markus Warzinsky, and participants
of the PDF4LHC meetings for useful discussions. 
This work was supported in part by the U.S.\ Department of Energy under
grant DE-FG02-04ER41299; by the Fermilab Research Alliance, LLC,
under Contract DE-AC02-76CH11359 with the United States Department of 
Energy; by the U.S. National Science Foundation under grant numbers 
PHY-0555545, PHY-0855561,
and PHY-0757758; by the National Science Council of Taiwan under
grant No. NSC-98-2112-M-133-002-MY3; by LHC Theory Initiative Travel
Fellowship awarded by the U.S. National Science Foundation under grant
PHY-0705862; and by Lightner-Sams Foundation. 
C.-P. Y. would also like to thank the hospitality of 
National Center for Theoretical Sciences in Taiwan and
Center for High Energy Physics, Peking University, in China, 
where part of this work was done.

\clearpage

%% file: pdf4app.tex
\appendix
\section{Production of a $W$-boson with $n$ partons in the LO-MC approach}

To study the impact of the LO-MC PDF sets on multi-parton configurations, of the kind commonly encountered
in parton shower-matrix element matching, we have performed a parton-level calculation
of $W^{+}+n$ parton cross sections ($n=0,1,2,3,4$) at the LHC center-of-mass energy of 10 TeV
using MADGRAPH~\cite{Alwall:2007st} and CTEQ6L1 and CT09MC2 PDF sets.
The final-state colored partons were required to have transverse momenta $k_T\ge 10$ GeV.  The
predicted cross  sections are presented in Table \ref{tab:mad}, broken down
into different subprocess components and ranked by the relative size.
For simplicity, we present only the results for up to two colored
partons in the final state, and we compute the ratio
$R_\sigma=\sigma(\mbox{CT09MC2})/\sigma(\mbox{CTEQ6L1})$ for each scattering channel. While
$R_\sigma$ is expected to vary between the different scattering channels, 
it is actually well represented by its average value in all channels,
$R_\sigma=1.26$. A notable exception is the $gg$ initial state, with $R_\sigma=1.48$.

\newlength{\locallinewidth}
\setlength{\locallinewidth}{\linewidth}

\begin{table}[ht]
\begin{center}
\begin{tabular}[c]{|p{0.2\locallinewidth}|p{0.15\locallinewidth}|p{0.15\locallinewidth}|p{0.1\locallinewidth}|} \hline \hline
Subprocess & $\sigma$(CTEQ6L1) (nb) & $\sigma$(CT09MC2) (nb) & Ratio
$R_\sigma$ \\ \hline \hline
\hline
$u\bar{d}{\to}e^{+}\nu_e$
 & 
7.323
 & 
9.029
 & 
1.23
 \\
\hline\hline

$ug{\to}e^{+}\nu_ed$
 & 
2.165
 & 
2.729
 & 
1.26
 \\
\hline

$u\bar{d}{\to}e^{+}\nu_eg$
 & 
1.760
 & 
2.207
 & 
1.25
 \\
\hline

$g\bar{d}{\to}e^{+}\nu_e\bar{u}$
 & 
0.835
 & 
1.130
 & 
1.35
 \\
\hline\hline

$ug{\to}e^{+}\nu_edg$
 & 
1.722
 & 
2.239
 & 
1.30
 \\
\hline

$g\bar{d}{\to}e^{+}\nu_e\bar{u}g$
 & 
0.546
 & 
0.751
 & 
1.38
 \\
\hline

$u\bar{d}{\to}e^{+}\nu_egg$
 & 
0.325
 & 
0.416
 & 
1.28
 \\
\hline

$gg{\to}e^{+}\nu_e\bar{u}d$
 & 
0.138
 & 
0.204
 & 
1.48
 \\
\hline

$uu{\to}e^{+}\nu_eud$
 & 
0.053
 & 
0.064
 & 
1.21
 \\
\hline

$ud{\to}e^{+}\nu_edd$
 & 
0.038
 & 
0.047
 & 
1.24
 \\
\hline

$u\bar{d}{\to}e^{+}\nu_ed\bar{d}$
 & 
0.028
 & 
0.036
 & 
1.29
 \\
\hline

$u\bar{d}{\to}e^{+}\nu_eu\bar{u}$
 & 
0.026
 & 
0.033
 & 
1.27
 \\
\hline

$u\bar{s}{\to}e^{+}\nu_ed\bar{s}$
 & 
0.022
 & 
0.027
 & 
1.23
 \\
\hline

$us{\to}e^{+}\nu_eds$
 & 
0.022
 & 
0.027
 & 
1.23
 \\
\hline

$u\bar{u}{\to}e^{+}\nu_e\bar{u}d$
 & 
0.020
 & 
0.024
 & 
1.20
 \\
\hline

$u\bar{d}{\to}e^{+}\nu_ec\bar{c}$
 & 
0.019
 & 
0.024
 & 
1.26
 \\
\hline

$c\bar{d}{\to}e^{+}\nu_e\bar{u}c$
 & 
0.015
 & 
0.019
 & 
1.27
 \\
\hline

$uc{\to}e^{+}\nu_eus$
 & 
0.013
 & 
0.016
 & 
1.23
 \\
\hline

$uc{\to}e^{+}\nu_ecd$
 & 
0.013
 & 
0.016
 & 
1.23
 \\
\hline

$d\bar{d}{\to}e^{+}\nu_e\bar{u}d$
 & 
0.012
 & 
0.016
 & 
1.33
 \\
\hline

$c\bar{u}{\to}e^{+}\nu_e\bar{u}s$
 & 
0.008
 & 
0.010
 & 
1.25
 \\
\hline

$u\bar{c}{\to}e^{+}\nu_e\bar{c}d$
 & 
0.007
 & 
0.009
 & 
1.29
 \\
\hline

$\bar{d}\bar{d}{\to}e^{+}\nu_e\bar{u}\bar{d}$
 & 
0.006
 & 
0.008
 & 
1.33
 \\
\hline

$u\bar{s}{\to}e^{+}\nu_eu\bar{c}$
 & 
0.006
 & 
0.008
 & 
1.33
 \\
\hline \hline
Total & 15.12 & 19.09  & 1.26 \\ \hline\hline
\end{tabular}
\end{center}
\caption{Breakdown of CTEQ6L1 and CT09MC2 cross sections and their
ratios for different subprocesses of $W^{+}+n \mbox{ jet}$ production ($n=0,\ 1,\ 2$) 
at the LHC center-of-mass energy of 10 TeV.}
\label{tab:mad}
\end{table}

A similar study for subprocesses containing three or four colored partons in the final
state reveals a similar pattern, but different values of $R_\sigma$. In these cases, 
$R_\sigma$ is equal to 1.34 for the total cross section, and $R_\sigma$ ranges 
from 1.48 to 1.77 for the gluon-gluon initial states.

The fact that $R_\sigma$ is different for different parton topologies will have
some phenomenological impact.  Color connections and parton types influence
the parton shower: gluons and quarks have different Sudakov form factors, and color
coherence limits the phase space for emissions. Thus, 
properties of jets resulting from a matched calculation
based on the CT09MC2 PDFs are likely to be different from those based
on CTEQ6L.  The scale of these differences remains to be seen.

\begin{figure}[ht]
\includegraphics[width=\textwidth]{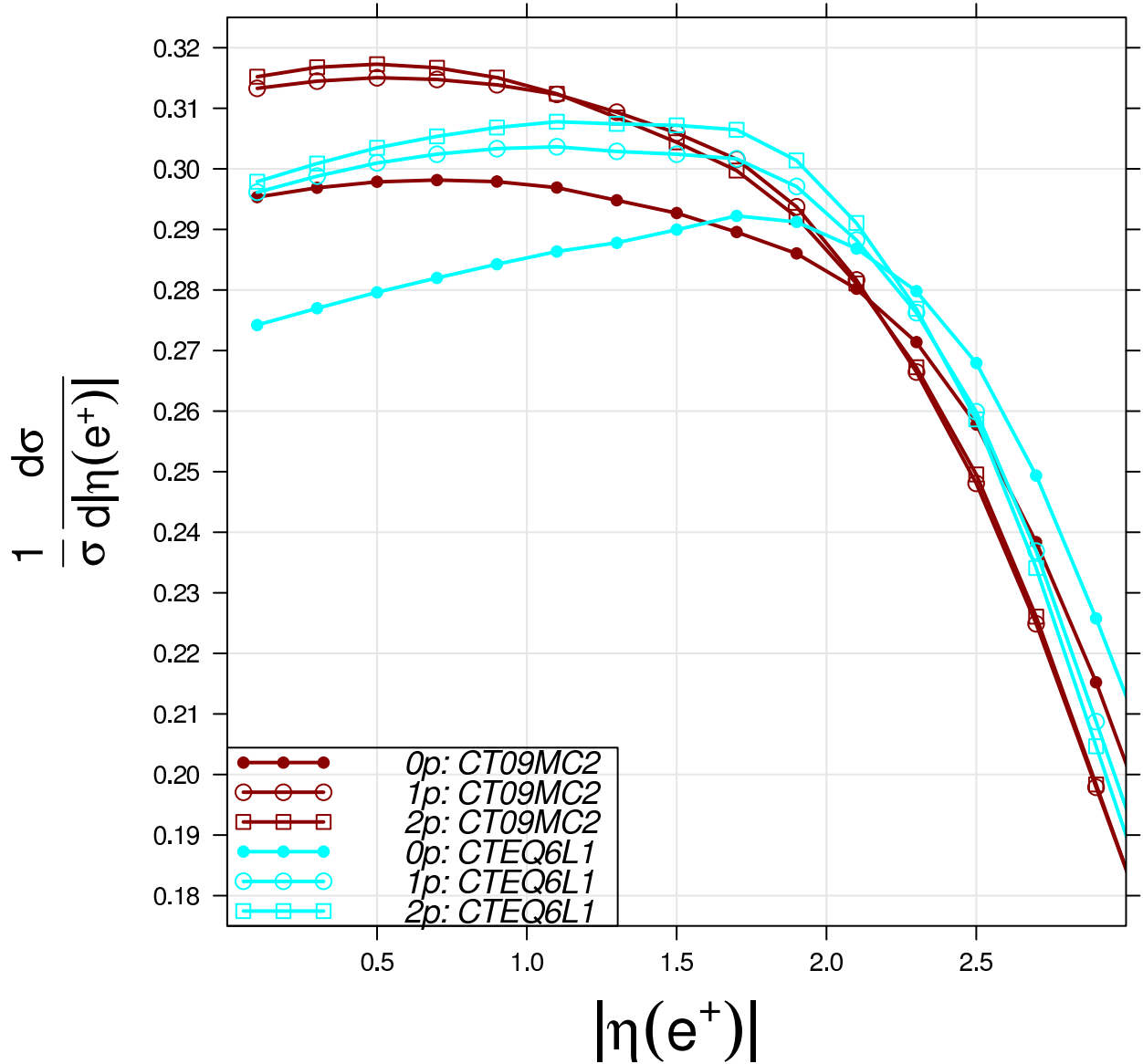}
\caption{Distribution of positron pseudorapidity $|\eta(e^{+})|$ for
various partonic $p$ multiplicities in the case of $W^{+}+n\mbox{ jets}$ ($n=0,1,2$) production
at the LHC (with a center of mass energy of 10 TeV), for CTEQ6L1 and CT09MC2 PDF sets.
The partonic jets are defined with $k_T\ge 10\mathrm{~GeV}$.
}
\label{fig:etaNp}
\end{figure}

The kinematics of partonic events is relatively unchanged between the
two PDF sets, except for the distributions of the particles from the 
$W^+$ boson decay.   
There is a tendency for colored partons to be more central with 
the CT09MC2 PDFs,
but the difference is not significant. The change in
shape of the rapidity distribution of the gauge boson has been
discussed in the main
text. The change in the distribution of the decay positron for
various partonic multiplicities is shown in Fig.~\ref{fig:etaNp}.
Two features are notable in this figure.   First, 
the CTEQ6L1 and CT09MC2 positron's rapidity distributions are different
for all parton multiplicities ($n=0,1,2$).   Second, for
a given PDF set, the distribution for $n=0$ is different from that 
for $n=1,2$. More detailed comparison will be deferred to future studies.

\clearpage

\section{K-factors for LO-MC PDFs}
The K-factor, calculated as the ratio of the NLO to LO cross sections,
depends on the choice of the renormalization/factorization scale,
the PDFs used, and the kinematic region being considered.
Even with the above caveats, it can be useful to define the K-factors
for physics processes at the LHC~\cite{Campbell:2006wx}.
Below we reproduce the table first shown in Ref.~\cite{Campbell:2006wx}
and then updated in the Les Houches 2007 proceedings,
where we have included the K-factors using our LO-MC PDFs
for processes at the LHC; 
a few of these processes were included as pseudodata in our global fit, while most were not.
The result is shown in Table \ref{tab:K-fact}.

In most cases, the K-factors are smaller (closer to unity)
when NLO PDFs are used for the LO calculations,
and this is true as well for predictions using the LO-MC PDFs.
The K-factor for $W$ production is less than 1 in this table:
the $W$ pseudodata used in the LO-MC fit were generated
with the CTEQ6.6M PDFs, which predict the LHC $W$ and $Z$ cross sections 
that are larger by 6-7\% than those based on CTEQ6 PDFs
(used as NLO cross sections in the K-factor table) \cite{Nadolsky:2008zw}.
In this way, the effects of the {\em variable flavor number heavy quark scheme}
used in the current NLO CTEQ PDF fits are effectively taken into account
in the LO-MC formalism.
The other quark-dominated process in the table below
(vector boson fusion  production of a 120 GeV Higgs)
also has a K-factor lower (0.75 compared to 0.85) when using the CTEQ6.6M PDFs
for the NLO calculation, rather than the CTEQ6 PDFs.
The K-factors for the other processes are nearly the same for CTEQ6M as for CTEQ6.6M. 

\begin{table}[h]
\begin{center}
\begin{tabular}{|l|l|l|c|c|c|c|c|c|c|}
\hline
  & \multicolumn{2}{|l|}{Fact. scales} & 
 \multicolumn{3}{|c|}{Tevatron K-factor} &
 \multicolumn{4}{|c|}{LHC K-factor} \\ 
  & \multicolumn{2}{|l|}{\quad} & 
 \multicolumn{3}{|c|}{} & \multicolumn{4}{|c|}{} \\
Process & $\mu_0$ & $\mu_1$ &
 ${\cal K}(\mu_0)$ & ${\cal K}(\mu_1)$ & ${\cal K}^\prime(\mu_0)$ &
 ${\cal K}(\mu_0)$ & ${\cal K}(\mu_1)$ & ${\cal K}^\prime(\mu_0)$ &
 ${\cal K}^{\prime\prime}(\mu_0)$  \\
\hline
&&&&&&&&&\\
$W$        & $m_W$ & $2m_W$	&
 1.33 & 1.31 & 1.21 & 1.15 & 1.05 & 1.15 & 0.95 \\
$W$+1 jet          & $m_W$ & $ p_T^{\rm jet}$ &
 1.42 & 1.20 & 1.43 & 1.21 & 1.32 & 1.42 & 0.99\\
$W$+2 jets & $m_W$ & $ p_T^{\rm jet}$	    &
 1.16 & 0.91 & 1.29 & 0.89 & 0.88 & 1.10 & 0.90\\
$WW$+1 jet~\cite{Dittmaier:2007th}~\cite{Campbell:2007ev}               & $m_W$ & $2m_W$    &
 1.19 & 1.37 & 1.26 & 1.33 & 1.40 & 1.42 & 1.10\\
$t{\bar t}$        & $m_t$ & $2m_t$         &
 1.08 & 1.31 & 1.24 & 1.40 & 1.59 & 1.19 & 1.09\\
$t{\bar t}$+1 jet~\cite{Dittmaier:2008uj}        & $m_t$ & $2m_t$    &
 1.13 & 1.43 & 1.37 & 0.97 & 1.29 & 1.10 & 0.85\\
$b{\bar b}$        & $m_b$ & $2m_b$         &
 1.20 & 1.21 & 2.10 & 0.98 & 0.84 & 2.51 & -- \\
Higgs      & $m_H$ & $ p_T^{\rm jet}$       & 
 2.33 & -- & 2.33 & 1.72 & -- & 2.32 & 1.43 \\
Higgs via VBF      & $m_H$ & $ p_T^{\rm jet}$ &
 1.07 & 0.97 & 1.07 & 1.23 & 1.34 & 0.85 & 0.83  \\
Higgs+1 jet     & $m_H$ & $ p_T^{\rm jet}$ &
 2.02 & 1.46 & 2.13 & 1.47 & 1.24 & 1.90 & 1.33\\
Higgs+2 jets~\cite{Campbell:2006xx}     & $m_H$ & $ p_T^{\rm jet}$
 & -- & -- & -- & 1.15 & -- & -- & 1.13 \\
&&&&&&&&&\\
\hline
\end{tabular}
\caption{\label{tab:K-fact}
K-factors for various processes at the LHC (at 14 TeV) calculated using
a selection of input parameters.
In all cases, for NLO calculations, the CTEQ6M PDF set is used.
For LO calculations, ${\cal K}$ uses the CTEQ6L1 set,
whilst ${\cal K}^\prime$ uses the same PDF set, CTEQ6M,
as at NLO, and ${\cal K}^{\prime\prime}$ uses
the LO-MC (2-loop) PDF set CT09MC2.
For Higgs+1 or 2 jets, a jet cut of $40\ \mathrm{GeV}/c$
and $|\eta|<4.5$ has been applied. 
A cut of $p_{T}^{\mathrm{jet}}>20\ \mathrm{GeV}/c$
has been applied to the $t\bar{t}$+jet process,
and a cut of $p_{T}^{\mathrm{jet}}>50\ \mathrm{GeV}/c$ to the $WW$+jet
process. In the $W$(Higgs)+2 jets process, the jets are separated by $\Delta R>0.4$ 
(with $R_{sep}=1.3$), whilst the vector boson fusion (VBF) 
calculations are performed for a Higgs boson of mass $120$~GeV.
In each case the value of the K-factor is compared
at two often-used scale choices, $\mu_0$ and $\mu_1$.}
\end{center}
\end{table}
